\DocumentMetadata{
  lang=en,
}	

\documentclass[sigconf,pdfusetitle,pdfdisplaydoctitle]{acmart}

\AtBeginDocument{%
  \providecommand\BibTeX{{%
    \normalfont B\kern-0.5em{\scshape i\kern-0.25em b}\kern-0.8em\TeX}}}

\usepackage{enumerate}
\usepackage{graphicx}
\usepackage{subcaption}
\usepackage{float}
\usepackage{multirow}
\usepackage{array}
\usepackage{makecell}
\usepackage{gensymb}
\usepackage{enumitem}
\usepackage[skip=3pt]{caption}  % Adjusts spacing above and below captions
\definecolor{lightgray}{gray}{0.95}  % Define a very light gray color

\copyrightyear{2026}
\acmYear{2026}
\setcopyright{cc}
\setcctype{by}
\acmConference[CHI '26]{Proceedings of the 2026 CHI Conference on Human Factors in Computing Systems}{April 13--17, 2026}{Barcelona, Spain}
\acmBooktitle{Proceedings of the 2026 CHI Conference on Human Factors in Computing Systems (CHI '26), April 13--17, 2026, Barcelona, Spain}
\acmPrice{}
\acmDOI{10.1145/3772318.3790589}
\acmISBN{979-8-4007-2278-3/2026/04}

\usepackage{xr-hyper}

\usepackage{accsupp}

\makeatletter
\newcommand*{\addFileDependency}[1]{% argument=file name and extension
  \typeout{(#1)}
  \@addtofilelist{#1}
  \IfFileExists{#1}{}{\typeout{No file #1.}}
}
\makeatother

\newcommand*{\myexternaldocument}[1]{%
    \externaldocument[supp-]{#1}%
    \addFileDependency{#1.tex}%
    \addFileDependency{#1.aux}%
}

\myexternaldocument{supplementary}

\definecolor{brownishred}{RGB}{0,0,0}
\DeclareRobustCommand{\colorchange}[1]{%
  \textcolor{brownishred}{#1}%
}

\newcommand{\system}{NaviNote}
\newcommand{\baseline}{TapTapSee}
\usepackage{xspace}
\usepackage{longtable}
\newcommand*{\eg}{e.g.,\@\xspace}
\newcommand*{\ie}{i.e.,\@\xspace}

\newcommand*{\vs}{vs.\@\xspace}
\newcommand*{\insitu}{in-situ\@\xspace}

\begin{document}

\title[\system{}]{\system{}: Enabling In-situ Spatial Annotation Authoring to Support Exploration and Navigation for Blind and Low Vision People}
\settopmatter{authorsperrow=4}
\author{Ruijia Chen}
\orcid{https://orcid.org/0000-0002-1655-6228}
\email{ruijia.chen@wisc.edu}
% \affiliation{\country{}}
\authornote{Authors contributed equally to this research. Order was determined alphabetically.}
% \\
% \textsuperscript{1}University of Wisconsin--Madison, Madison, Wisconsin, USA.\\
% \textsuperscript{2}Niantic Spatial, Inc., London, United Kingdom\\
% \textsuperscript{3}Niantic Spatial, Inc., Zaragoza, Spain\\
% \textsuperscript{4}University College London, London, United Kingdom}
\affiliation{%
  \institution{University of Wisconsin-Madison}
  \city{Madison}
  \state{Wisconsin}
  \country{USA}
}
% \affiliation{
%   \textsuperscript{1}University of Wisconsin-Madison， Madison, Wisconsin, USA\\
%   \textsuperscript{2}Niantic Spatial, Inc., London, United Kingdom\\
%   \textsuperscript{3}University College London, London, United Kingdom
% }
\author{Yuheng Wu}
\orcid{0009-0005-1828-400X}
\email{yuheng.wu@wisc.edu}
\authornotemark[1]
 % \affiliation{\country{}}
\affiliation{%
  \institution{University of Wisconsin-Madison}
  \city{Madison}
  \state{Wisconsin}
  \country{USA}
}

\author{Charlie Houseago}
\orcid{0000-0001-8700-5397}
\email{charlie@nianticspatial.com}
 % \affiliation{\country{}}
\affiliation{%
  \institution{Niantic Spatial, Inc.}
  \city{London}
  \country{United Kingdom}
}

\author{Filipe Gaspar}
\orcid{0000-0001-5363-4961}
\email{filipe@nianticspatial.com}
 % \affiliation{\country{}}
\affiliation{%
  \institution{Niantic Spatial, Inc.}
  \city{London}
  \country{United Kingdom}
}

\author{Filippo Aleotti}
\orcid{0000-0002-8911-3241}
\email{filippo@nianticspatial.com}
 % \affiliation{\country{}}
\affiliation{%
  \institution{Niantic Spatial, Inc.}
  \city{London}
  \country{United Kingdom}
}

\author{Dorian Gálvez-López}
\orcid{0009-0009-0463-2279}
\email{dorian@nianticspatial.com}
 % \affiliation{\country{}}
\affiliation{%
  \institution{Niantic Spatial, Inc.}
  \city{Zaragoza}
  \country{Spain}
}

\author{Oliver Johnston}
\orcid{0009-0002-3423-3526}
\email{ojohnston@nianticspatial.com}
 % \affiliation{\country{}}
\affiliation{%
  \institution{Niantic Spatial, Inc.}
  \city{London}
  \country{United Kingdom}
}

\author{Diego Mazala}
\orcid{0000-0002-7481-802X}
\email{diego@nianticspatial.com}
 % \affiliation{\country{}}
\affiliation{%
  \institution{Niantic Spatial, Inc.}
  \city{London}
  \country{United Kingdom}
}

\author{Guillermo Garcia-Hernando}
\orcid{0000-0003-3215-7857}
\email{guillermo@nianticspatial.com}
 % \affiliation{\country{}}
\affiliation{%
  \institution{Niantic Spatial, Inc.}
  \city{London}
  \country{United Kingdom}
}

\author{Maryam Bandukda}
\orcid{0000-0002-2367-6471}
\email{m.bandukda@ucl.ac.uk}
 % \affiliation{\country{}}
\affiliation{%
  \institution{University College London}
  \city{London}
  \country{United Kingdom}
}

\author{Gabriel Brostow}
\orcid{0000-0001-8472-3828}
\email{brostow@cs.ucl.ac.uk}
 % \affiliation{\country{}}
\affiliation{%
  \institution{University College London}
  \city{London}
  \country{United Kingdom}
}

\author{Jessica Van Brummelen}
\orcid{0000-0002-4831-6296}
\email{jess@nianticspatial.com}
\affiliation{%
  \institution{Niantic Spatial, Inc.}
  \city{London}
  \country{United Kingdom}
}
 % \affiliation{\country{}}

\renewcommand{\shortauthors}{Chen and Wu, et al.}

\begin{abstract}
GPS and smartphones enable users to place location-based annotations, capturing rich environmental context. Previous research demonstrates that blind and low vision (BLV) people can use annotations to explore unfamiliar areas. However, current commercial systems allowing BLV users to create annotations have never been evaluated, and current GPS-based systems can deviate several meters. Motivated by high-accuracy visual positioning technology, we first conducted a formative study with 24 BLV participants to envision a more accurate and inclusive annotation system. Surprisingly, many participants viewed the high-accuracy technology not just as an annotation system but also as a tool for precise last-few-meters navigation. Guided by participant feedback, we developed \system{}, which combines vision-based high-precision localization with an agentic architecture to enable voice-based annotation authoring and navigation. Evaluating \system{} with 18 BLV participants showed that it significantly improved navigation performance and supported users in understanding and annotating their surroundings. Based on these findings, we discuss design considerations for future accessible annotation authoring systems.
\end{abstract}

\begin{CCSXML}
<ccs2012>
   <concept>
       <concept_id>10003120.10011738.10011775</concept_id>
       <concept_desc>Human-centered computing~Accessibility technologies</concept_desc>
       <concept_significance>500</concept_significance>
       </concept>
   <concept>
       <concept_id>10003120.10011738.10011776</concept_id>
       <concept_desc>Human-centered computing~Accessibility systems and tools</concept_desc>
       <concept_significance>500</concept_significance>
       </concept>
   <concept>
       <concept_id>10003120.10011738.10011774</concept_id>
       <concept_desc>Human-centered computing~Accessibility design and evaluation methods</concept_desc>
       <concept_significance>500</concept_significance>
       </concept>
   <concept>
       <concept_id>10003120.10003121.10003124.10010392</concept_id>
       <concept_desc>Human-centered computing~Mixed / augmented reality</concept_desc>
       <concept_significance>500</concept_significance>
       </concept>
 </ccs2012>
\end{CCSXML}

\ccsdesc[500]{Human-centered computing~Accessibility technologies}
\ccsdesc[500]{Human-centered computing~Accessibility systems and tools}
\ccsdesc[500]{Human-centered computing~Accessibility design and evaluation methods}
\ccsdesc[500]{Human-centered computing~Mixed / augmented reality}

\keywords{Accessibility, Blind and Low Vision, Last-Few-Meters Navigation, Spatial Annotation, Augmented Reality, Visual Positioning System, Large Language Models}

%TC:ignore

{
\begin{teaserfigure}
        \centering
\vspace{-1.5em}
        \includegraphics[width=\textwidth]{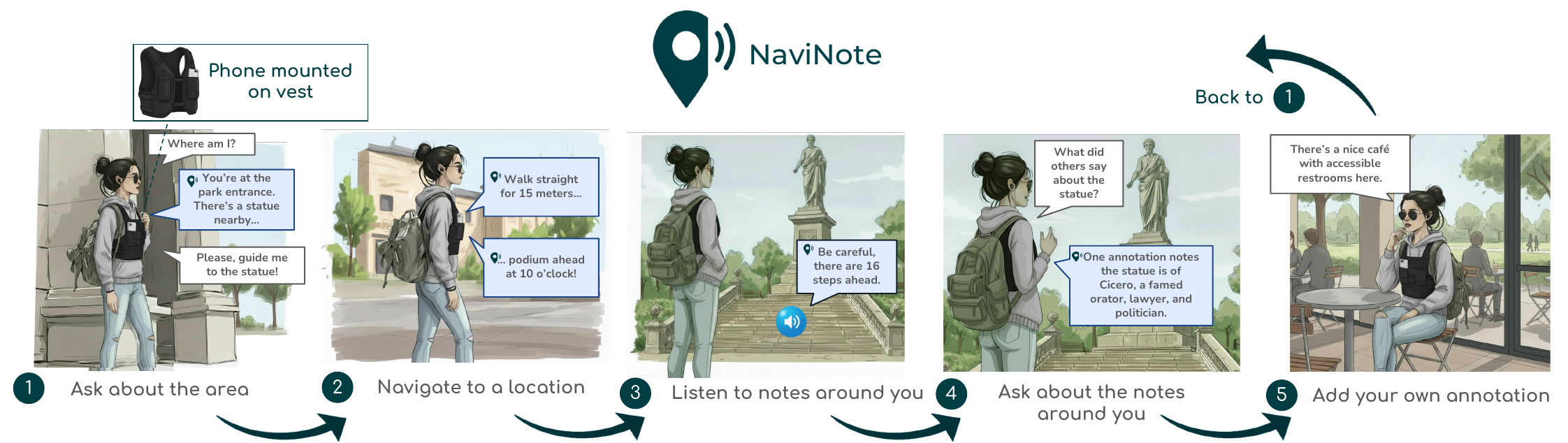}
  \caption{
  \system{} enables blind and low vision (BLV) users to explore their surroundings via a five-stage pipeline \colorchange{after localizing their precise positions using Visual Positioning System (VPS) in a pre-scanned area on a smartphone}: (1) User asks about the area using natural language; (2) User navigates to a specific location within the area by following the turn-by-turn instructions \system{} provides; (3) User listens to nearby spatial annotations created by other BLV users; (4) User asks follow-up questions on spatial annotations and from other information sources; (5) User creates their own spatial annotations. \colorchange{Arrows in the figure indicate the internal flow of the pipeline. During the interaction, the user wears a vest that holds the smartphone with camera facing forward.}}
 \Description{A five-stage pipeline is shown with 5 panels. Each panel shows a user with a vest and speech bubbles communicating between \system{} and the user. In (1), the user asks, ``Where am I?'' The system responds: ``You're at the park entrance. There's a statue nearby...'' The user follows up: ``Please, guide me to the statue!''; In (2) \system{} provides navigation instructions, ``Walk straight for 15 meters...'', ``... podium ahead at 10 o'clock''; In (3) the user listens to a nearby spatial annotation, ``Be careful, there are 16 steps ahead''; In (4) the user asks: ``What did others say about the statue?'' The system responds: ``One annotation notes the statue is of Cicero, a famed orator, lawyer and politician.''; In (5) the user creates their own spatial annotation, ``There's a nice café with accessible restrooms here.''. During the interaction, the user wears a vest that holds the smartphone with camera facing forward, and localizes their precise positions using Visual Positioning System (VPS) in a pre-scanned area on the smartphone.}
  \label{fig:teaser}
\end{teaserfigure}
}
\maketitle

\section{Introduction}
Exploring and understanding one's surroundings is essential for human well-being \cite{Russell2013Humans, Finlay2019Closure} yet remains a persistent challenge for blind and low vision (BLV) people.
Large language models (LLM) and computer vision (CV) capabilities have supported BLV users with \insitu exploration by detecting nearby objects \cite{islam2024identifying, meshram2019astute, morrison2023understanding, lin2025ai}, recognizing text \cite{huang2019augmented, prabha2022vivoice}, and providing scene descriptions \cite{Penuela2024Investigating, hao2024multi, magay2024light, hao2024chatmap, seeingaiSeeingTalking, rao2021google, hao2024multi, kuribayashi2025wanderguide},
yet these camera-based AI tools impose limitations for BLV users around aiming the camera correctly \cite{coughlan2020audio,chang2024worldscribe}, verifying information, and addressing LLM hallucinations (\ie models generate plausible-sounding but unfaithful or nonsensical information \cite{ji2023towards}) \cite{alharbi2024misfitting,herskovitz2023hacking,yu2024human}. They also cannot provide information about objects outside the camera's frame.

Location-based annotations complement existing tools by providing verified contextual information outside the camera view (\eg an entrance behind the user). Prior work, Footnotes \cite{gleason2018footnotes}, has shown their value in providing multifaceted information to BLV users that LLM- and CV-based tools alone cannot provide.
However, BLV users play a passive role in FootNotes as well as other assistive systems for exploration \cite{kaniwa2024chitchatguide, coughlan2020towards, froehlich2025streetviewai}, consuming information along predetermined routes rather than contributing their own situated knowledge.
Existing systems are also limited by the precision of localization technologies, often relying on the 10-meter accuracy of GPS \cite{merry2019smartphonegpsaccuracy} or similar, which limits the utility of annotations as they cannot be precisely placed or accessed. 

New emerging technologies could bridge this gap, allowing the creation of accessible, empowering interfaces for BLV users to act as active \insitu explorers and annotation contributors. First, multi-modal large language models (MLLM) can enable voice-first interfaces for users to interact accessibly with their surroundings \cite{kaniwa2024chitchatguide}. Second, the latest Visual Positioning Systems (VPS) are far more precise than GPS at tracking where the user's phone is in 3D space \cite{lightshipVisualPositioning, horvath2025googlevpsaccuracy}, without needing to aim cameras at specific targets. We propose that together, these technologies offer opportunities for high-accuracy, conversational, \insitu spatial annotations. We therefore seek to answer the following research questions in the context of such a system:
\begin{itemize}
    \item RQ1: What types of annotations are most valuable to BLV users, and in what scenarios are they most applicable?
    \item RQ2: How can we design an accessible interface to enable BLV users to author \insitu spatial annotations?
    \item RQ3: How can BLV users effectively access and utilize spatial annotations?
\end{itemize}

To answer these questions, we first conducted a formative study involving 24 BLV participants, identifying six key annotation categories: \textit{Safety}, \textit{Accessibility}, \textit{Amenity}, \textit{Layout}, \textit{Attraction}, and \textit{Experience}. Beyond annotations, we learned our proposed technology could address a critical prerequisite for exploration: precise ``last-few-meters'' navigation, \colorchange{providing turn-by-turn instructions during the final meters of travel where traditional GPS often fails \cite{saha2019closing}}. Motivated by these findings, we designed \textit{\system{}}, a voice-based system combining VPS technology with an AI-powered conversational interface. Utilizing sub-meter-level localization (\ie position error of less than one meter), \system{} provides turn-by-turn instructions and an audio compass to facilitate safe navigation for BLV users, and supports \insitu accessing and authoring of annotations at precise locations.

To evaluate \system{}, we conducted a second study with 18 BLV participants in a local public square, now focusing on navigation and free exploration. Sixteen participants compared navigation using \system{} versus a baseline, and all 18 participants completed a free-form exploration \colorchange{to listen to pre-set annotations and create new ones}. 
We found \system{} significantly improved navigation performance, with 14 of 16 participants successfully navigating using \system{} compared to only six with the baseline\colorchange{, and also enabled all participants to successfully create annotations independently.} Moreover, \system{} significantly outperformed the baseline across qualitative measures, including perceived effectiveness, usability, mental demand, frustration, and perceived performance. Based on participants' \insitu authoring, we refined the taxonomy of desired annotations and introduced another category, \textit{Request}. 

In summary, our contributions include: (1) \system{}, a voice-based system that enables BLV users to create \insitu spatial annotations and navigate effectively in last-few-meters scenarios, (2) an evaluation of \system{} with 18 BLV participants, demonstrating its benefits for navigation and exploration, and (3) a taxonomy of BLV users' desired spatial annotations, developed through two studies ($n_1$=24, $n_2$=18). 
\section{Related Work}
We situate our system’s initial design within prior work on supporting BLV users in understanding their surroundings, navigating the last-few-meters, and using \insitu spatial annotations, as well as on crowdsourcing for accessibility.

\subsection{User-Initiated Tools for Environmental Understanding}
Assistive technologies often help BLV users understand their surroundings through user-initiated requests. Many of these are camera-based, supporting magnification \cite{wittich2018effectiveness, stearns2018design}, object recognition \cite{islam2024identifying, meshram2019astute, morrison2023understanding}, and text recognition \cite{huang2019augmented, prabha2022vivoice} in real time. However, properly aiming a camera is often difficult for BLV users \cite{hong2024understanding, jayant2011supporting, vazquez2012helping}, and the information returned is typically limited to describing \colorchange{each individual object} or reading text without providing broader contextual cues needed for navigation and orientation \cite{herskovitz2023hacking}.

Several tools have introduced conversational AI to improve BLV users' access to environmental information. Seeing AI \cite{seeingaiSeeingTalking} and Be My Eyes \cite{bemyeyes2015}, two widely used smartphone apps \cite{gamage2023blind}, let users capture photos and receive detailed audio descriptions. Research prototypes push this further: ChitChatGuide \cite{kaniwa2024chitchatguide} supports question answering about nearby points of interest (POIs) using LLMs; WorldScribe \cite{chang2024worldscribe} generates real-time visual descriptions tailored to users' contexts (\eg concise descriptions for dynamic scenes, detailed descriptions for stable settings); and StreetViewAI \cite{froehlich2025streetviewai} enables virtual exploration of destinations via LLM-accessible streetscape maps. These systems move beyond recognition to enable richer, context-sensitive queries. Yet they still primarily position BLV users as passive consumers: while users can ask questions and receive descriptions, they cannot contribute new, \insitu knowledge to the location.

To our knowledge, no tools currently enable BLV users to share situated knowledge and lived experiences about the physical world \insitu. Our research addresses this gap.

\subsection{In-Situ Spatial Annotations and ``Last-Few-Meters'' Navigation}
Spatial annotations bind information to locations, providing answers beyond camera-based recognition. Prior work has described this as creating ``chatty environments,'' where the environment ``speaks'' to users to provide otherwise inaccessible information \cite{coroama2003chatty}. For sighted users, spatial annotations have long been studied. For example, GeoNotes \cite{persson2003geonotes} enabled users to attach notes to specific locations using PDAs, ActiveCampus \cite{griswold2004activecampus} allowed students to share annotations to discover labs and nearby events, and StoryPlace.Me \cite{bentley2012storyplace} explored how elderly users could leave location-based stories for future generations. %Similarly, Google Maps reviews \cite{google_maps_reviews_help} can be seen as a form of \exsitu spatial annotation.

A few spatial annotation systems have targeted BLV users \cite{may2020spotlights, gleason2018footnotes, coughlan2020audio,coughlan2017ar4vi,shen2013camio,coughlan2020towards}, with FootNotes \cite{gleason2018footnotes} being most relevant. In FootNotes, researchers attached annotations to existing OpenStreetMap objects, which would play when BLV participants' GPS signals were nearby. Yet, the system did not provide an interface designed specifically for BLV users to author annotations, and rather assumed screen readers and text-to-speech would suffice without user evaluation. \colorchange{Several GPS-based commercial systems, such as Sendero BrailleNote GPS \cite{senderogroupSenderoBrailleNote}, Trekker Breeze \cite{afbMadeSimple}, BlindSquare \cite{blindsquareBlindSquare}, and Microsoft Soundscape \cite{microsoftMicrosoftSoundscape}, allowed BLV users to actively mark their environments. Users can search for and learn about important nearby POIs (\eg crossroads, stores, popular cafes), mark their own POIs, and save routes for later retracing via voice interaction. However, these GPS-based systems, including FootNotes and commercial tools, face the key constraint of the ``last-few-meters'' problem \cite{saha2019closing, gleason2018footnotes, fiannaca2014headlock, lock2017portable}, where GPS fail to provide accurate turn-by-turn guidance in the final few meters---sometimes as big as tens of meters \cite{modsching2006field}---due to signal inaccuracy or incomplete map data \cite{saha2019closing}. As a result, BLV users fail to accurately locate and reach the intended POIs. Even when BLV users know they are near or directly facing a POI, they often lack adequate information (\eg haptic textures, smells, or characteristic sounds for landmark recognition \cite{saha2019closing, loomis2001navigating}) to confidently use the location with GPS.}

To bridge this gap, some systems have introduced infrastructure for higher accuracy, such as Bluetooth beacons (\eg NavCog3 \cite{sato2019navcog3}), NFC‑tagged landmarks \cite{ganz2014percept}, or ultrawide-bandwidth beacons \cite{lu2021assistive}, but these require costly hardware installations and pre-built floor maps. Others have explored navigation without pre-built maps \cite{yoon2019leveraging,kuribayashi2023pathfinder,kuribayashi2025wanderguide, saha2019closing}. For example, WanderGuide \cite{kuribayashi2025wanderguide} allowed users to explore the environment with a suitcase-shaped robot, providing users with walkable directions and customizable levels of audio scene descriptions. Clew \cite{yoon2019leveraging} enabled users to record and later retrace routes using smartphone sensors. Closing the Gap \cite{saha2019closing} demonstrated the potential of camera-based landmark recognition to assist users near a destination, but relied on a Wizard-of-Oz setup \cite{dahlback1993wizard} rather than a deployable pipeline. None of these systems provided BLV users with a low-cost, low-barrier-to-entry, independent solution to the last-few-meters problem.

\colorchange{Visual Positioning Systems (VPS) provide a new opportunity for the last-few-meters navigation at high accuracy. In addition to GPS information, VPS combine input from multiple sensors (\eg camera, motion sensors, depth sensors) to generate visual features in the surrounding environment. As users move their phone and scan nearby objects or buildings, VPS compare what the camera sees with a large database of previously mapped visual features to pinpoint the user’s exact location and orientation with sub-meter accuracy \cite{lightshipVisualPositioning, rajpurohit2023review, lee2020semantic, google2025geospatial}.
Motivated by the gaps in accessible annotation authoring and precise last-few-meters navigation, \system{} tackles these dual challenges with natural language interaction and VPS–based localization to empower BLV users to navigate to, access, and author their own spatial annotations independently using a smartphone.}

\subsection{Crowdsourcing for BLV Users}
Crowdsourcing annotations has been widely used to make locations more accessible for BLV users. Previous research explored online crowdsourcing to annotate bus stops \cite{prasain2011stopfinder, hara2015improving}, accessible sidewalks \cite{hara2013combining, saha2019project, li2024never}, storefronts \cite{liu2022annotating}, obstacles \cite{rice2013crowdsourcing}, semantic information in the real world environment \cite{gleason2016vizmap, guo2018crowd}, and alt text for online images \cite{salisbury2017toward}. Other work demonstrated how crowdsourced descriptions of artwork can enhance museum experiences for BLV users \cite{petrie2023crowdsourcing}.

In addition to large-scale labeling, several systems connect BLV users with sighted volunteers for on-demand assistance. For example, Be My Eyes \cite{bemyeyes2015} allows BLV users to share their camera view with sighted helpers for real-time support, while VizWiz::LocateIt \cite{bigham2010vizwiz} enables BLV users to post photos and questions online, allowing remote workers to outline objects and help them locate targets. 

Despite these advances, most crowdsourcing tools cast BLV users as requesters rather than independent contributors. Yet BLV users hold valuable perspectives, which, when shared, can better address their communities’ needs \cite{herskovitz2023hacking}. As AI becomes increasingly capable of automatic annotation, it is critical to identify what BLV users value most, rather than relying on what AI ``decides'' is important. Our two studies investigate these values and needs, and \system{} empowers BLV users to act as active information contributors.

\section{Formative Study (Study 1)}
To inform the design of \system{}, we conducted a formative study with 24 BLV participants, focusing on whether they would value highly-localized audio annotations, what they might use them for, and where they might place them (RQ1). The study identified BLV participants' desired types of annotations and revealed how precise navigation would be a prerequisite for annotation authoring.

\subsection{Methods}

\subsubsection{Participants ($n_1$=24)}
We recruited 24 BLV participants (F1-F24) from the local BLV communities. Table \ref{tab:demographics_formative} in Appendix \ref{sec:formative_demographics} lists participant demographics and visual conditions. Participants received a compensation of 75 GBP.
% equivalent to \colorchange{100 USD} in GBP.

\subsubsection{Procedure}
We conducted a user-centric design workshop, which consisted of two identical sessions over two days with 12 participants each day. In each session, participants were randomly assigned to one of three four-person groups. Each session included four stages: scenario-based brainstorming, tactile-map prototyping, Wizard-of-Oz (WoZ) authoring \cite{dahlback1993wizard}, and reflection. Both sessions lasted for approximately three hours.

\textbf{Scenario-based Brainstorming.} Inspired by the co-design activity in \cite{brewer2018facilitating}, we introduced participants to seven different scenarios (see Supplementary Section 2%\ref{supp-sec:formative-scenario-design-space}
), where real-world spatial annotations with centimeter-level accuracy could be useful. %, and had a short discussion about how the participants could imagine this working. 
Participants then voted and held small-group discussions on the top two scenarios, which were as follows. \textbf{Day 1}: (1) What audio annotations could Orientation \& Mobility (O\&M) teachers add to a space to act as a training ground for you? (2) You're at home organizing your room. What annotations would you place to help with organization? \textbf{Day 2}: (1) You're traveling to a friend's apartment complex to house-sit for them. What annotations would you want them (or the complex) to place ahead of time? (2) You're making a location-based audio game. What annotations would you place?

\begin{figure*}[htb]
    \centering
    \includegraphics[width=\textwidth]{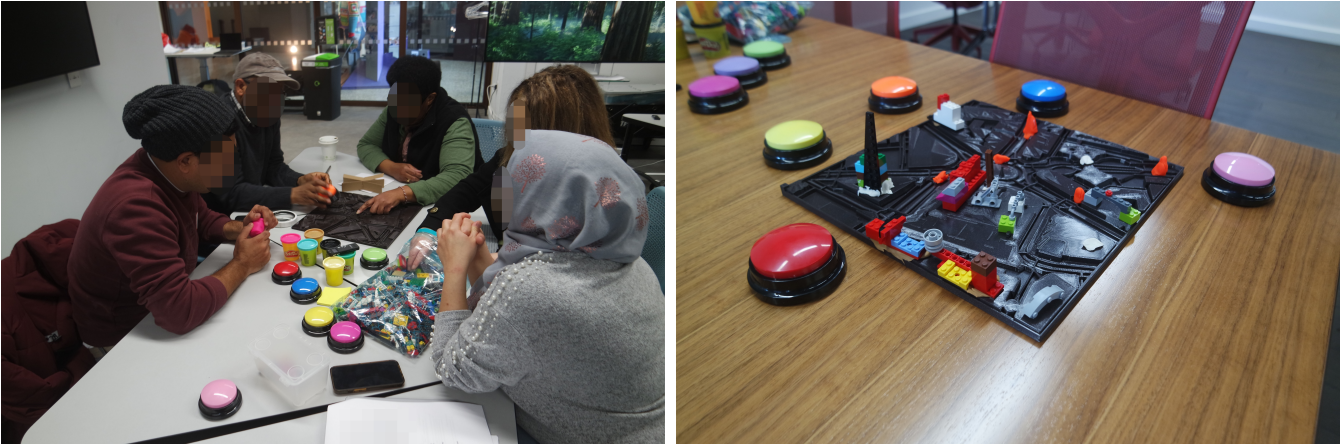}
    \caption{Left: One of the formative study small groups prototyping where they would place annotations on a 3D-printed map. Right: An example of a final prototype, with various audio buttons---which contain annotation recordings---arranged around a tactile map. The orange modeling clay indicates annotation locations.}
    \Description{The first image shows five people sitting at a table with a tactile map. One participant is touching the map, likely identifying objects. Others are holding modeling clay. The second image shows five large physical buttons arranged around a tactile map. Various bits of modeling clay and Lego pieces are arranged on the map.}
    \label{fig:formative-prototyping}
\end{figure*}

\textbf{Tactile-Map Prototyping.} We provided each group with a 3D-printed tactile map of a well-known local area, \eg a historical site, a large intersection, or a park with a garden. Participants audio-recorded annotations, and placed markers on the map indicating the annotation locations (see Figure \ref{fig:formative-prototyping}). They prototyped for approximately 45 minutes and described their prototypes to the larger group.

%TC:ignore
%TC:endignore

\textbf{WoZ Authoring.} We conducted a %non-deceptive 
WoZ study on UCL East campus, where one researcher acted as an AI system that participants could ask questions of---\eg about the surroundings or to place an audio note---and another researcher acted as the location-based audio playback system using an audio recorder. Each group engaged with the WoZ separately for 15 minutes, authoring their own annotations indoors and outdoors on campus.

\textbf{Reflection.} Participants engaged in semi-structured discussion for 15-30 minutes in a large group about the pain points of the proposed system, how they envisioned the system's form factor, and the most and least important system features. (See Supplementary Section 3 %\ref{supp-sec:formative-discussion-questions}
for full list of discussion questions.)

\subsection{Analysis}
\label{subsec:formative_analysis}
We audio-recorded all study sessions, transcribed them using Whisper \cite{radford2022whisper}, and manually fixed transcription errors. Transcripts were analyzed using thematic analysis \cite{braun2006using,clarke2017thematic}: Two researchers open-coded the transcripts from the first day of the workshop (approximately 50\% of the data) independently, and developed a codebook by discussing their codes to resolve disagreements. The two researchers then split the transcripts from the second day, coded independently based on the initial codebook, and added new codes when necessary. During this process, the two researchers periodically discussed new codes to ensure consistency. New codes were added to the codebook upon agreement.
The research team then developed themes and sub-themes from the codes by clustering relevant codes using axial coding and affinity diagrams \cite{terry2017thematic}. Finally, the researchers cross-referenced the original data, the codebook, and the themes to make final adjustments, ensuring all codes were correctly categorized. For this formative analysis, our emphasis is on the range of perspectives rather than numerical prevalence \cite{maxwell2010numbersinqualitative}. We present them as motivation and grounding for our system design.

\subsection{Findings}
Here, we summarize participants' motivation for spatial annotations and present six categories of annotations they desired. We also present their preferred system features to derive design guidelines for \system{}.

\subsubsection{Annotation Motivation}
Participants identified two primary purposes for spatial annotations. First, they hoped to use annotations to share and access information within their social circles (\eg facilities in their home when their friends were visiting) and the broader BLV community (\eg safety warnings for other BLV people). Second, participants envisioned documenting \insitu experiences, either for personal use (\eg revisiting and recollecting memories) or for collective benefit (\eg sharing comments and feedback about the accessibility of locations).

Beyond location-based annotations, participants also described potential needs for dynamic object-based annotations, \colorchange{primarily for indoor scenarios,} such as locating and organizing household items (\eg matching clothing, identifying spices or cleaning products), leaving operational instructions or reminders of food expiration dates for safety, and attaching personal memories to specific objects like photos. While equally important, we focus on location-based annotations in this work and leave dynamic object annotations for future exploration.

\subsubsection{Annotation Taxonomy}
\label{subsubsec:taxonomy}
We identified six categories of desired \colorchange{spatial} annotations for BLV users: safety, accessibility, amenity, layout, attraction, and experience. We noticed these categories often overlapped with those from FootNotes \cite{gleason2018footnotes}, but also extended them---\eg our ``Attraction'' category broadens FootNotes' narrower focus on ``Historical''---reflecting the wider range of annotations found in participant quotes.

\textbf{\textit{Safety.}} Participants emphasized safety-related annotations, including depth-related (\eg stairs, curbs, uneven pavement, cracks) and width-related features (\eg narrow entrances). They were attentive to height, width, orientation, and number of steps due to challenges in depth perception. Furthermore, they wanted annotations near water (\eg rivers, fountains, low-fenced areas) to alert them of risks of falling, and annotations about protruding or sharp obstacles such as hanging branches and thorns. Participants also highlighted crossroads as important to be annotated, \colorchange{including the position of zebra crossings, curbs again, and ``where the traffic lights (buttons) are.'' (F2)}

\textbf{\textit{Accessibility.}} Participants noted a need for annotations about accessible facilities (\eg elevators, escalators, ramps). For example, they proposed annotations distinguishing accessible toilets \colorchange{(\ie toilets specially designed to accommodate people with disabilities, such as those equipped with braille signage and offering larger space)} from generic ones to help find them faster. Participants also wanted annotations that identified ``haptic paths,'' \colorchange{as F15 described, ``something with a touch and tail.'' They wanted to annotate features that could be held or followed with a white cane, such as handrails, walls, or curbs.}
Participants also wanted to annotate help points (\eg reception desks and concierges). As F2 noted: ``You may need to know where the reception desk is, so you can speak to someone.''

\textbf{\textit{Amenity.}} 
Participants wanted annotations about amenities, including public transportation (\eg train stations, bus stops), common functional areas (\eg shops, parks, cafes, prayer rooms), and facilities (\eg toilets, bins, benches). They were also interested in information related to such amenities, such as opening and closing times, public access, and peak hours of use.

\textbf{\textit{Layout.}}
Consistent with prior work \cite{chen2025visimark}, besides annotations on individual locations, participants also desired an overview of their surroundings, such as the overall shape and size of their current area, locations of entrances and exits, and the start and end of passageways. They also wanted information about the boundaries of different areas, such as the boundary between pedestrian zones and bike lanes.

\textbf{\textit{Attraction.}}
Participants wanted to annotate tourist attractions, such as providing historical details about a location, similar to QR code-based tours in museums. They also wanted annotations that shared information about upcoming events, such as exhibitions nearby, and envisioned annotating salient landmarks (\eg a big tree) to support mental map construction and to share as meeting points.

\textbf{\textit{Experience.}}
Finally, participants wanted to create annotations documenting personal experiences and comments about particular places. As F3 noted: ``It would be nice to have a note, like, oh, `I came here with my friend, and we went to this building.' '' Participants also suggested that such experience annotations could serve as a community feedback channel, similar to ``an audio version of Google Maps reviews'' (F4) \cite{GoogleMaps2025}, where they can provide suggestions \colorchange{to service providers for improving} accessibility. \colorchange{For example, F23 suggested that cafes should offer assistance to people with additional needs when making purchases or payments upon entering the store.}

\subsubsection{Design Guidelines}
\label{subsubsec:desired_features}
Participants shared their desired features for an annotation-authoring system. We summarize their needs into six design guidelines (DG1-6) for \system{}. Several of these guidelines also align with findings from prior work \cite{froehlich2025streetviewai}.

\textbf{\textit{DG1: Support Navigation with Constant Confirmation.}} Participants across groups in the Reflection activity agreed that a wayfinding feature would be essential: Why have annotations if you can't independently explore the world? 

\textbf{\textit{DG2: Support Different Levels of Conversational Queries.}} Participants highlighted how the system should answer questions about objects in front of them, and provide higher-level information about their surroundings. 
This could include recommendations about nearby places or concise summaries of the area. As F23 detailed, ``I was thinking, `What if they have something all in one, like a kind of a introduction of anywhere, and it will tell you what you expect. What's in front of you? When you turn there?' It's like a two minute [introduction].''

\textbf{\textit{DG3: Play Certain Annotations Automatically, Especially Safety.}} Participants suggested some annotations should be played automatically, noting that even seemingly irrelevant information might turn out to be useful. As F24 said, ``We should never fathom that it was there, because if you don't switch it on, you don't know, right? If you just point at something you want and you don't realize what's around you.'' All participants noted safety annotations as most important and that they should be played automatically, possibly alongside vibrations or beeping alerts.

\textbf{\textit{DG4: Allow for Information Filtering.}} 
While agreeing that certain annotations should be played actively, participants also expressed wanting annotation filters, % to avoid sensory overload. 
which could be customized based on visual abilities and preferences. %, and environmental conditions. 
As F2 explained: ``If I'm in an area that I'm fairly familiar with, I may need less [information] layers to make sure I'm walking straight. Whereas if I'm in a completely new area that is unfamiliar to me, I may want a bit more information.'' F7 added, ``During the day, I actually don't need that much information. And then when it suddenly goes dark... I need more information. Then it's like, `I nearly tripped.' ''

\textbf{\textit{DG5: Enable BLV Accessibility and Customizability.}}\label{DG5: Accessible and Customizable for BLV.} Participants stressed that the system must be accessible and adaptable to diverse needs\colorchange{, suggesting the use of voice input and tactile buttons. In the Tactile-Map Prototyping phase, large physical buttons (Figure~\ref{fig:formative-prototyping}, right) were used to simulate recording audio annotation, inspiring participants to recommend similar tactile buttons alongside voice input for more accessible control in the final design.}

Given VPS requires \colorchange{camera information}, participants suggested attaching it to the body to simplify aiming. %, or providing guidance to help aim the camera at targets accurately. 
Moreover, they highlighted the importance of customization, such as adjusting voice settings, and navigation directions (\eg clock-face \vs left-and-right), %, and modes of transport (\eg by bus or on foot), 
which aligns with prior findings on the need for personalization \cite{herskovitz2023hacking, chang2024worldscribe, froehlich2025streetviewai}.

\textbf{\textit{DG6: Offer Multi-modal Feedback.}} Participants expected the system to offer both haptic and audio feedback. For example, vibrations could be used for obstacle warnings, while spatial audio could guide users toward destinations. Visual feedback was considered important for low-vision users, as they relied heavily on residual visuals \cite{szpiro2016finding}. Interestingly, blind participants also wanted visual feedback for sighted companions who might assist them. As F7 explained, ``I'm surrounded with people with full vision. And sometimes when it's all audio, they're like, `I can't help you'. So I feel like for the sighted people that are going to assist me, they do need a visual cue.''

\section{\system{}: Voice-based Annotation and Navigation System for BLV Users}

Based on the formative study findings, we developed an interaction framework to enable BLV users to access spatial annotations independently (RQ2). We then designed and implemented \system{}, the first system that enables BLV users to conversationally create and access \insitu{} annotations. \system{} is hands-free and voice-based, and supports querying the surroundings, navigating to objects or annotations of interest, and creating, editing and deleting annotations. We included customization features to allow users to tailor AI responses and UI to their needs. %We detail our system design and implementation below.

\subsection{Interaction Pipeline for Annotation Authoring}
% \subsection{System Design and Implementation}
\label{subsec:system_design}
%Based on the formative study findings, we propose an 
Our five-stage interaction pipeline for accessible exploration and annotation authoring follows:

\textit{\textbf{Scanning}}. \system{} relies on pre-scanned and pre-processed environment data from a Point of Interest (POI) (\ie a geographic area of interest \cite{SUN2023poi}) to accurately determine users' pose (\ie position, orientation) and understand scene elements. Relying on pre-scanned data removes the burden for BLV users to properly scan the environment in real time \cite{lee2025imaginatear} or aim the camera precisely \cite{chang2024worldscribe,alharbi2024misfitting}. Various other research and industry applications utilize scans in the same way, including ImaginateAR \cite{lee2025imaginatear}, CoCreatAR \cite{numan2025cocreatar}, Google Geospatial API \cite{google2025geospatial}, and Niantic Wayfarer for Pokémon GO \cite{niantic2025wayfarer}, and have shown that crowdsourcing scans can effectively gather wide, accurate coverage \cite{nummenmaa2025employing, niantic2025crowdsource}. We envision this pre-scanning process being performed by sighted users, volunteers \cite{niantic2025wayfarer}, or even other BLV users \cite{dontlooknow}. \colorchange{We discuss the implementation details of pre-scanning and pre-processing in Section \ref{subsubsec:scanning}.}

\textit{\textbf{Localization}}. Upon launching \system{}, users perform a brief localization process (5-10 seconds) by slowly moving their devices horizontally and vertically to observe the surrounding environment. The system detects visual features and to determine the user's pose with sub-meter-level accuracy. Once localized, \system{} fetches the associated spatial data for navigation and annotation.

\textit{\textbf{Query}}. Users can ask \system{} questions about their surroundings or other topics of interest (\eg history of a nearby statue). \system{} provides responses by synthesizing information from multiple sources, including the pre-scanned environment, existing spatial annotations, users' camera feed, public map sources, and the internet.

\textit{\textbf{Navigation}}. %Users can navigate to a particular object or annotation to further explore the area. 
When asked, \system{} provides turn-by-turn instructions to guide users to target objects or annotations, and provides hazard warnings en route drawn from safety-related annotations.

\textit{\textbf{Annotation Authoring}}. Users can create annotations at specific locations by conversing with the system, such as by saying ``Place an annotation on this bench saying it has armrests.'' They can also edit or remove existing annotations via a similar conversational procedure.

\subsection{System Design and Implementation}
Based on the interaction pipeline and formative study, we designed \system{} with four major components: Scanning and Localization, Query Answering, Navigation, and Annotation Accessing and Authoring.
\system{} contains a phone-based frontend for user interaction and a Python server backend for request processing.
We built the frontend with Unity 6000.0.34f1 and Niantic SDK for Unity \cite{Niantic2025ARDK}, and deployed the prototype on an iPhone 14 Pro with iOS 18.6.1.  
Regarding the form factor, we used a commercial running vest with adhesive straps to secure the phone on the users' chest for hands-free operation (\eg while carrying a white cane). For audio output, we used L5 Bone Conduction Earphones, as bone conduction earphones enable users to hear system instructions while still perceiving environmental sounds simultaneously~\cite{asakura2021bone}. Wired earphones were also provided as an alternative.

%TC:ignore
\begin{figure*}[htb]
    \centering
    \includegraphics[width=\textwidth]{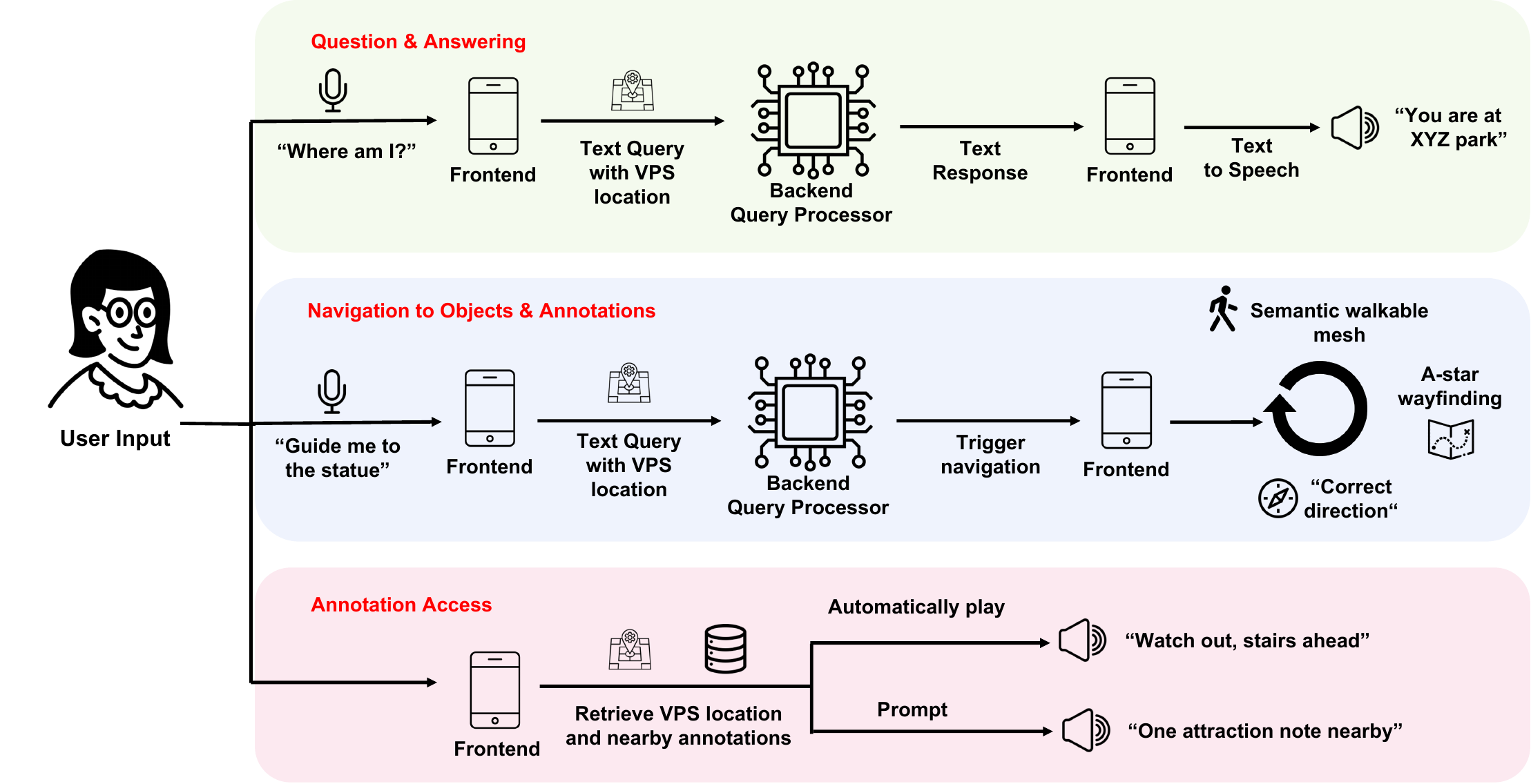}
    \caption{\colorchange{The interaction flow of \system{}}. \textbf{Top}: \colorchange{Question answering}. The user asks a question to the phone, which converts speech to text and sends the query with VPS location to the backend. The backend returns a text response, which the frontend speaks aloud via text-to-speech. \textbf{Middle}: \colorchange{Navigation to objects and annotations of interest}. The user requests guidance to an object or annotation. The query and location are sent to the backend, which triggers navigation on the frontend. As the user moves, the frontend generates turn-by-turn instructions with an audio compass \colorchange{with A-star wayfinding algorithm} over the semantic walkable mesh. \textbf{Bottom}: \colorchange{Annotation access}. The frontend continuously retrieves VPS location and nearby annotations, playing them either automatically or by prompt.}
     \Description{A diagram that starts with ``User Input'' and branches in three directions. The first branch shows the Question and Answering pipeline, beginning with ``Where am I?'', which is provided to a phone (frontend) and converted to a text query. This text query with VPS location is sent to a backend ``Query Processor'', which develops a text response. The response is sent to the frontend and converted to speech, ``You are at XYZ park.'' The second branch shows the Navigation to Objects and Annotations during the navigation pipeline. The user states, ``Guide me to the statue,'' which goes through the same process as in the Question and Answering pipeline, but instead of just providing a text response, the query triggers a navigation cycle. This cycle runs on the frontend and includes generating a semantic walkable mesh, \colorchange{using A-star wayfinding algorithm to generate instruction}, and providing user feedback, \eg by saying ``Correct direction.'' Finally, the third branch shows the Annotation Access pipeline. The backend retrieves the VPS location and nearby annotations, which then either automatically play or are prompted. The ``automatically play'' branch says, ``Watch out, stairs ahead,'' and the ``prompt'' branch says, ``One attraction note nearby.''}
    \label{fig:data_flow}
\end{figure*}
%TC:endignore

\subsubsection{Scanning and Localization}
\label{subsubsec:scanning}
For \system{} to dynamically track users' pose and understand their surrounding environment, we first scan the whole POI to generate its 3D map.
We adopt Niantic's VPS framework \cite{lightshipVisualPositioning} both when making and using our bespoke \system{} map, \colorchange{as this framework supports crowdsourced scans from commercial devices with depth sensors (\eg recent iPhone or Android phone equipped with LiDAR) and provides high coverage of popular POIs \cite{nianticlabsBuildingLarge}. Users conduct the scanning by pointing their devices at the surroundings, while the framework captures synchronized multi-modal sensor streams including visual imagery, depth information, and device poses. In our experiment setup, scanning the study area of approximately $45$ m $\times$ $45$ m takes $22$ minutes. These streams are aggregated and uploaded into Niantic's VPS server where the data is transformed into spatial features for localization in future sessions.}

Besides spatial features, we also construct a \textit{scene graph} of each POI from its 3D map. \colorchange{A scene graph describes the spatial properties (\ie position, rotation, and dimensions, represented as bounding boxes in the 3D point cloud) of each object within the POI. Similar to the scanning process, the scene graph can be generated through crowdsourcing by sighted users and volunteers.} Prior work also demonstrated the practicality of automatically generating scene graphs from RGB-D inputs \cite{gu2023conceptgraphsopenvocabulary3dscene, wei20233d, Lv2023SGFormer:}. In our user evaluation study, we manually annotated the scene graph of the experimental scene to ensure accuracy.
% and avoid the need for users to validate competing scene graph generation methods.
% bypass the need for users to validate competing points-to-scene-graph systems.

Ultimately, when a user is running \system{}, the system loops, estimating the user's current pose relative to the scene graph and \colorchange{fetching} spatial annotations from the backend, and providing these as context to \system{}'s conversational agent system.

\subsubsection{Query Answering}
One of \system{}'s most notable features is its capability to answer user queries with natural language, as illustrated in the top panel of Figure \ref{fig:data_flow}. This was motivated by DG2 and implemented 
%Driven by the advancements and widespread adoption of agentic AI systems,  we designed the backend of \system{} using MLLMs. % to enable natural language interaction for accessibility.
%The backend is structured 
in the ``Orchestrator'' pattern \cite{openai_agent} with MLLM agents. The backend contains a single entry point (\ie the orchestrator) for user queries and has access to modules specialized for particular purposes (\ie agents). The orchestrator analyzes user queries, delegates them to agents with suitable tools to gather relevant information, and synthesizes a final response.

We implemented six agents\colorchange{, each designed to handle different types of user queries, supporting real-time understanding of the surroundings through the camera and external databases (\eg our pre-defined annotations and public map data).} \textbf{Annotation Agent}, which retrieves, creates, edits, and removes spatial annotations. \colorchange{It answers questions related to existing annotations (\eg ``What do people think of this park?'')}; \textbf{Scene Understanding Agent}, which interprets scene graphs of POIs and computes object positions to answer spatial queries (\eg ``What's in front of me?''), and triggers navigation on demand; \textbf{Image Search Agent}, which accesses users' camera frames and answers visual questions (\eg ``What color is this flower?''\colorchange{, ``Please read the sign.''}); \textbf{Internet Search Agent}, which retrieves information from the internet; \textbf{Public Map Agent}, which uses public map data (\eg Open Street Map \cite{OpenStreetMap}) to gather geographic information of nearby areas at a larger scale with coarser granularity (\eg nearest train station) to supplement the scene graphs; and \textbf{Customization Agent}, which interprets user customization requests and sends them to the frontend for execution. All agents have access to users' pose provided by VPS from the frontend. We used OpenAI GPT-4o \cite{OpenAI2025GPT4o} for the orchestrator and GPT-4o mini \cite{OpenAI2025GPT4oMini} for the agents.
% The orchestrator and agents are powered by models from the GPT-4 family, with the specific model variant being selected depending on the complexity of each Agent.

The backend is hosted on a cloud machine and communicates with the frontend via WebSockets \cite{fette2011websocket}. Users query by speaking to the frontend, which transcribes queries locally using Whisper \cite{radford2022whisper} and forwards the transcribed text to the orchestrator. The orchestrator invokes relevant agents and triggers actions on the frontend as needed, and finally aggregates agents' outputs into a response, which is converted to audio on the frontend via iOS text-to-speech \colorchange{API}. The orchestrator is instructed to give responses within two sentences or approximately 45 words by default, and users can customize its verbosity. Users can also provide their visual conditions or desired information (\eg ``include color information'') to help the orchestrator produce more helpful responses. We provide the system prompt in Supplementary Section 5.%\ref{supp-sec:system-prompt}.

Following participants’ preference for physical buttons (DG5), we reprogrammed the physical volume button to serve as the interaction trigger. To issue a query, the user presses the volume button \colorchange{(either on the smartphone or on connected earphones)}, speaks their question, and waits for the system to process the input.

\subsubsection{Navigation}
\label{subsubsec:navigation}
\system{} supports last-few-meters navigation to facilitate scene exploration, as illustrated in the middle panel of Figure \ref{fig:data_flow}. As the scene graph only contains information about static objects within the POI and does not reflect the environment's dynamics (\eg people nearby), \system{} uses the Semantic Mesh provided by Niantic SDK \cite{Niantic2025SemanticMesh} to construct a real-time mesh of walkable ground. We further refine the walkable mesh by excluding parts that collide with objects in the scene graph. % to reduce potential AI misrecognition. 
Then, \system{} computes the shortest path from the user to the destination using the A-star algorithm \cite{hart1968formal}. The path is then smoothed to retain only key turning points. If the path aligns with the edges of objects in the scene graph, it forms a ``haptic path'' (Section \ref{subsubsec:taxonomy}) and generates navigation instructions accordingly (\eg ``Correct direction. Follow the edge of the flower bed on your right'').

\begin{figure}[htb]
    \centering
    \includegraphics[width=.45\textwidth]{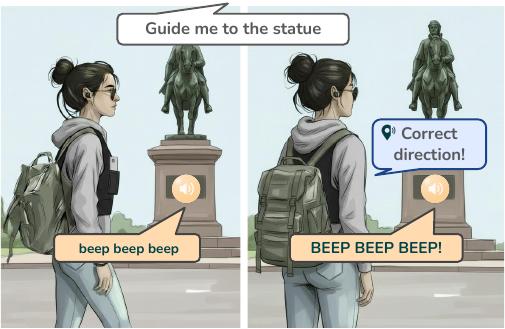}
    \caption{During navigation, there is an audio signal (``beep'') in the direction users should be moving. When the user's orientation is correct, the audio signal continuously plays at high volume, whereas at a lower volume with an incorrect orientation. When the user's orientation is correct, the system also periodically provides verbal confirmation, saying, ``Correct direction!''}
    \Description{The image depicts a user wearing a vest and earphones with a phone attached in front of her. During navigation, there is a spatial audio signal (``beep'') in the direction users should be moving. For example, when guiding a user to a statue, a continuous high-volume beep (``BEEP'') indicates correct orientation, whereas a lower volume indicates an incorrect orientation. When the user faces the statue correctly, the system also periodically provides verbal confirmation, saying ``Correct direction!''}
    \label{fig:audio_compass}
\end{figure}

\begin{figure*}[htb]
    \centering
    \includegraphics[width=\textwidth]{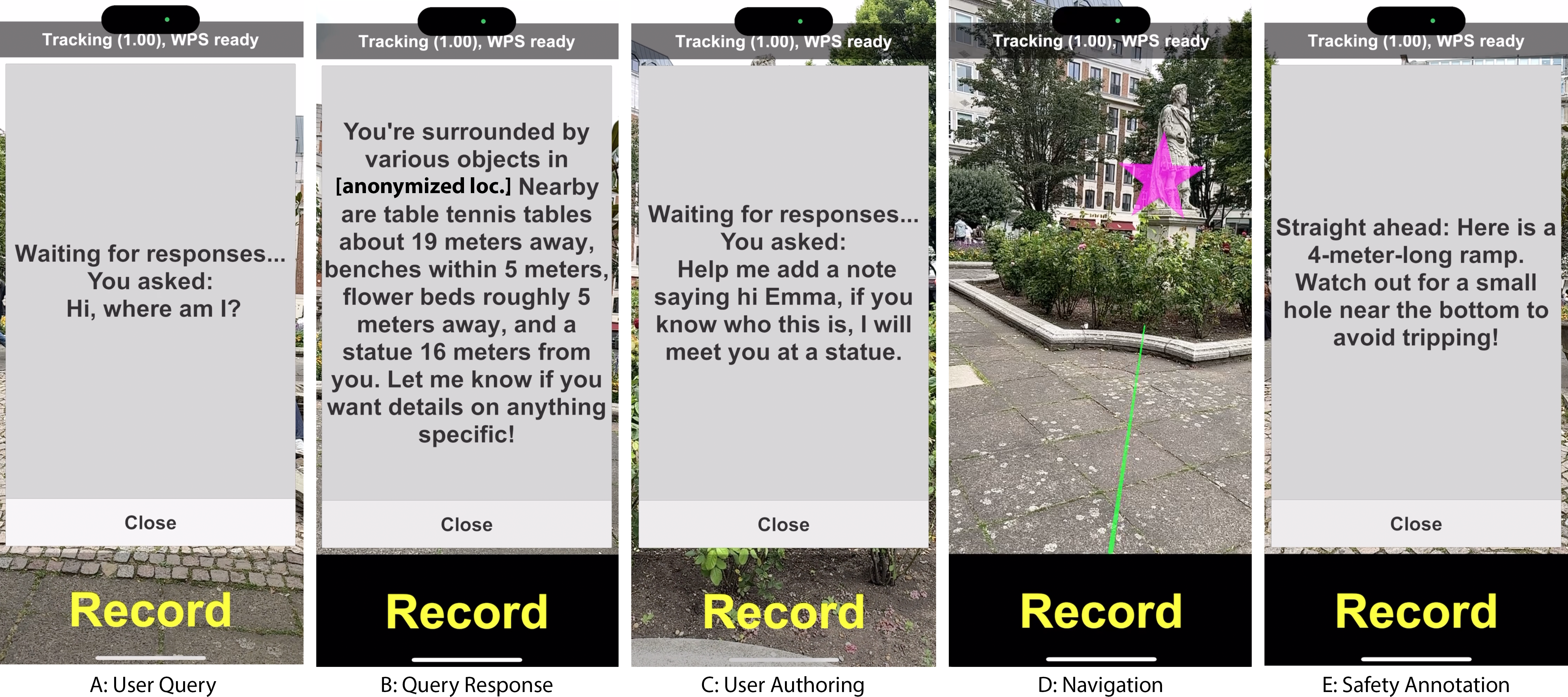}
    \caption{\system{}'s high-contrast visual interface. In (A), the user has asked where they are, and is waiting for a response from the system. In (B), the system responds. In (C), the user asks the system to place an annotation. In (D), the system shows a path to the statue during navigation. In (E), a safety annotation automatically plays. Note that the interface is voice-first, and this visual interface serves to address DG6.}
     \Description{Five panels showing \system{}'s visual interface. The interface has a large panel with text either showing what the user asked (\eg ``Hi, where am I?'' (A) and ``Help me add a note saying, `Hi Emma, if you know who this is, I will meet you at a statue.'  '' (C) ), the system's response (\eg ``You're surrounded by various objects in [a local public square]. Nearby are table tennis tables about 19 meters away, benches within 5 meters, ...'' (B) ), or annotations (\eg ``Straight ahead: Here is a 4-meter-long ramp. Watch out for a small hole near the bottom to avoid tripping!'' (E) ). The one exception is panel (D), which shows the navigation interface: A bright green line pointing towards a bright pink star, overlaid on the real-world camera view. The pink star sits at the location of a statue.}
    \label{fig:navinote_interface}
\end{figure*}

\system{} delivers navigation instructions in two ways: turn-by-turn instructions and audio compass. Turn-by-turn instructions are announced in a concise format (\eg ``2 o'clock, 5 meters''). To provide timely feedback of direction (DG1), they are triggered at turning points and every 10 seconds as confirmation (\eg ``Correct direction, 2 meters''), and when the user deviates from the planned route. The distance to the destination is also updated regularly (\eg ``Destination lies at 12 o'clock''). All audio instructions are played in spatial audio to provide directional cues \cite{katz2011spatial,katz2012navig}. Additionally, we developed an ``audio compass'' to provide users with immediate and constant feedback on their direction (DG1). The audio compass is a beeping signal that continuously plays at high volume when the user is facing the correct direction and at low volume when not (see Figure \ref{fig:audio_compass}). Following DG6, we also implemented visual feedback, including a green-line path and a star marker at the destination (see Figure \ref{fig:navinote_interface}).%, to allow users with residual vision to see the instructions.

%TC:ignore
%TC:endignore

\system{} allows users to customize navigation instructions. They can choose among different directional formats, including clock-face (\eg ``Turn to 11 o'clock''), egocentric (\eg ``Turn forward-right''), egocentric with degrees (\eg ``Turn left by 30 degrees''), and cardinal (\eg ``Facing south. Head East''). Users can also select their preferred unit of distance: meters, feet, or steps (with one step defined as 0.76 meters).

\subsubsection{Annotation Accessing and Authoring}
\label{subsubsec:annotation_authoring}
\system{} supports accessing existing spatial annotations (see the bottom panel of Figure \ref{fig:data_flow}) and authoring new ones. Spatial annotations are triggered when the user approaches within a predefined distance ($1.5$ meters for safety and accessibility annotations, and $1.0$ meter for other less critical categories). 
Following DG3 and DG4, \system{} provides two mechanisms for accessing annotations \colorchange{to provide layered access to avoid information overload. Different access mechanisms are applied based on annotation category: by default,} \textit{Safety} and \textit{accessibility} annotations are \textbf{automatically played} to ensure timely delivery of critical information. For safety annotations, the frontend also vibrates for 0.5 seconds (DG6). % to alert users of potential hazards. 
The other four categories are accessed \textbf{only by asking}: when nearby, the system delivers a prompt with a short sound effect (``bing'') followed by a notification aggregating these annotations (\eg ``two attraction notes nearby''). Users can ask follow-up questions to access them. 
\colorchange{When multiple annotations are nearby, annotations set to play automatically are prioritized over those set to be prompted. All messages are played sequentially, and any message that cannot be played within 15 seconds is skipped to avoid information overload.} Users can also adjust which categories of annotations are automatically played, prompted, or ignored by talking with the system (see Section \ref{subsec:apparatus} for participants' choice).

For authoring, users can create annotations in natural language by simply describing the new annotation (\eg ``Place an annotation here saying, `Exit ahead'.'') Annotations can be placed on objects in the area or at the user's current location. The backend automatically determines the annotation category from the content and acknowledges successful creation. Users can also modify or remove annotations they previously authored through follow-up commands (\eg ``I want to delete the annotation about the exit''). \colorchange{We set all annotations to be publicly accessible for the convenience of prototyping, and leave the management of public and private annotations as future work (see more discussion in Section 5 %\ref{subsubsec:discussion_privacy}
).}

%TC:ignore
%TC:endignore

\section{Evaluation (Study 2)}
To evaluate the effectiveness of \system{} in assisting BLV people in exploring outdoor environments (RQ3), we conducted a study with four pilot participants and 18 study participants in a public local square. We aim to answer the following questions: (1) What did BLV users ask \system{} during outdoor exploration?, (2) How effective is \system{} in supporting BLV users in last-few-meters navigation?, (3) What do BLV users think about the exploration features, including the process of annotation access and authoring, and the existing annotation types?, %(4) What do BLV users think about the system's interaction design?, and (5) 
and (4) How do BLV users envision using \system{} in other real-world scenarios?

\subsection{Participants (\texorpdfstring{$n_2$=18}{n2=18})}
\label{subsec:evaluation_participants}
% We recruited 4 BLV participants for a pilot study (2 females, 2 males). All of them participated in the formative workshop: Pilot1 (F7), Pilot2 (F19), Pilot3 (F5), Pilot4 (F23).
We recruited 18 BLV participants (11 females, 7 males) from local low-vision communities and via snowball sampling \cite{parker2019snowball}. Their ages ranged from 20 to 80 ($Mean = 48.9$, $SD = 16.5$). Six participants reported being totally blind and eight reported being legally blind. %Eight participants had prior experience with AR applications. 
Table \ref{tab:demographics_evaluation} in the Appendix \ref{sec:evaluation_demographics} details participants' demographics and visual conditions. Eight participants (P1-P4, P6, P10, P12, P16) also took part in the formative study. Two participants did not complete the navigation task (Section \ref{subsec:procedure}): P17 withdrew by choice and P18 was interrupted externally. Both participants only experienced \system{} during tutorial and the free exploration task. All 18 participants received compensation of 75 GBP.
% equivalent to \colorchange{100 USD} in local currency.%, including travel reimbursement. 

\colorchange{We also recruited four BLV participants (see Table \ref{tab:demographics_evaluation} in the Appendix) for pilot studies before the formal evaluation. After these pilot studies, we refined the experimental scripts to better explain system features and gathered valuable feedback from participants that informed our baseline condition choice (Section \ref{subsec:apparatus}). We also fixed system engineering bugs, and adjusted the annotation trigger distances from 2 meters to 1.5 meters for safety and accessibility annotations, and to 1.0 meter for other categories (Section \ref{subsubsec:annotation_authoring}), to help users better locate annotations while still receiving safety alerts in advance.}

\subsection{Apparatus} \label{subsec:apparatus}

\begin{figure*}[htb]
    \centering
    \includegraphics[width=\textwidth]{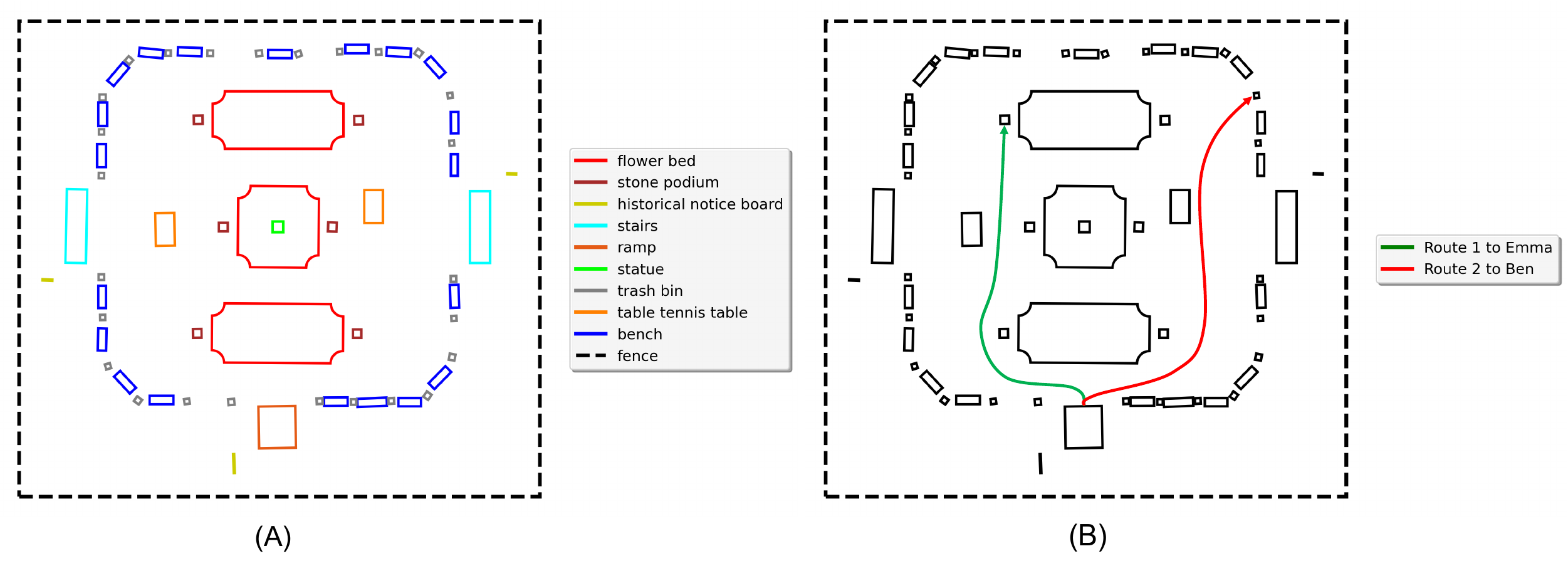}
    \caption{(A) The layout of the local public square for the evaluation study (gates omitted). There are 20 benches, 26 trash bins, two set of stairs, one ramp, six stone podiums, three flower beds, two table tennis tables, three historical notice boards, and one statue. (B) The two routes for the navigation task. Participants start from the ramp and navigate to the furthest podium on the left for ``Emma'' (green route), and to a trash bin on the right back corner of the square for ``Ben'' (red route).}
    \Description{
    (A) Map of the fenced square for the evaluation study. The square consists of various objects: three large flower beds in the center, two table tennis tables, two set of stairs, 20 benches arranged along the edges, six stone podiums near the flower beds, 26 trash bins by the benches, three historical notice boards at west, east, and southern side of the square, a ramp at the southern end, and a statue at the center of the central flower bed.
    (B) An illustration of the two routes for the navigation task. Route 1 to Emma starts from the ramp, curves around the left side of the central flower bed, and ends near the top left of the square at a stone podium. Route 2 to Ben also begins at the ramp, goes upward along the right side of the central flower bed, and ends near the top right of the square at a trash bin. The two routes are of similar length.
  }
    \label{fig:square-layout-route}
\end{figure*}

We conducted the study in a local public square (Golden Square in London) selected for its size (\(\sim \)$40\times40$ m), which represented a realistic navigation scenario, and for its safety, with fencing along the perimeter. \colorchange{The ground surface within the experimental area was uniformly flat and did not contain any designated tactually salient pathways, though some existing structures (\eg curbs, flower beds) were naturally perceivable by touch.} See Figure \ref{fig:square-layout-route}(A) for the square layout. We planned two routes for participants to navigate, each approximately 30 m long with an equal number of turns (\ie two turns) \cite{ishikawa2008wayfinding, may2020spotlights} and a similar number of landmarks (\eg table tennis tables, flower beds) \cite{yesiltepe2021landmarks,millonig2007developing}, as shown in Figure \ref{fig:square-layout-route}(B). \colorchange{Since the square contains three large elevated flower beds with curbs in the middle, both designed routes required participants to walk around one of the flower beds.} The research team created 39 annotations within the square, with 11 safety, 7 accessibility, 4 amenity, 6 layout and 4 experience annotations, which were all inspired by quotes and findings from the formative study, as shown in Supplementary Section 6.%\ref{supp-sec:annotations_inspired_by_quotes}.

Given the lack of existing tools to address the last-few-meters problem for BLV people, we used photo describer tools (\eg SeeingAI \cite{seeingaiSeeingTalking}) as baselines for navigation. Initially, we tested the Apple Magnifier \cite{AppleMagnifier} in a pilot study with four BLV participants. Participants reported that it caused information overload, due to its continuous stream of information. Pilot4 commented: ``This is something too much to keep listening and listening,'' and Pilot1 added: ``Because too much [information caused] information [over]load, you might think it's not [even] information.'' Pilot4 suggested \baseline{} \cite{TapTapSee2025} for the baseline, which we adopted, as it is commonly used in the BLV community. \baseline{} delivers concise, on-demand image descriptions by double tapping the screen. We considered audio-based tools as baseline, such as VoiceVista \cite{voicevista2025}, which offers audio beacons for navigation. However, the beacons rely on GPS localization, which is not precise enough for last-few-meters navigation \cite{merry2019smartphonegpsaccuracy}.

%TC:ignore
%TC:endignore

\subsection{Procedure}
\label{subsec:procedure}
The study consisted of a single on-site session that lasted approximately 2.5 hours. Prior to the on-site session, participants completed a demographics questionnaire (see Appendix \ref{sec:evaluation_demographics}). The on-site session contained \colorchange{four} parts: a tutorial on \system{}, a navigation task, a free-form exploration, \colorchange{and an exit interview}.
%about their demographic information, visual conditions, and prior experiences with AR and navigation technologies.

\subsubsection{Tutorial} 
We started with a tutorial on \system{} and the navigation baseline, \baseline{}. For \system{}, the localization process was completed in advance by the research team to ensure consistent localization quality across all participants, allowing the later evaluation to begin from the same starting point. We first instructed participants to press the volume button of the earphone to ask questions. Fifteen participants used bone conduction earphones and three (P1, P3, P4) opted for wired alternatives. We demonstrated how to create, edit, and delete annotations using natural language, and how to access annotations either automatically or on demand. Participants also learned about the navigation function, including turn-by-turn instructions and the audio compass, through a short practice route that did not overlap with the later experiment. \colorchange{We ensured that both the high and low audio compass volumes were audible and comfortable for all participants.} Participants were welcome to customize voice timbre (\eg gender, accent) and speed, and set which type of annotations they wanted to be played automatically, prompted, or ignored. \colorchange{All participants kept \textit{safety} and \textit{accessibility} annotations set to play automatically, with the remaining categories set to be prompted.} For \baseline{}, we explained how to double tap the screen to take a picture and wait 7–10 seconds for the image description \cite{TapTapSee2025}. Participants practiced using both systems freely until fully confident about their usage.

\subsubsection{Navigation Task}
We contextualized the navigation task in the scenario of meeting two friends, named ``Emma'' and ``Ben,'' at specific locations in the square. We provided initial destination descriptions before the task (\eg ``Emma told you she'd meet you at the podium with the vase, which is the leftmost podium furthest from you if you’re starting from the ramp'') and marked the destination with large text labels. For \system{}, an annotation was placed at the two destinations (Supplementary Section 6%\ref{supp-sec:annotations_inspired_by_quotes})
, and participants could ask the system to provide turn-by-turn instructions to the meeting spots. \colorchange{Although \system{} supports creating annotations during navigation, participants were asked to explore this feature later during the free-form exploration to ensure consistent comparison of navigation time.} For \baseline{}, participants could use it to get scene descriptions as needed during navigation. Researchers followed the participants during the walk to ensure safety, and participants could stop the task at any time. We counterbalanced the order of the two systems (\system{} or \baseline{}) and the two routes using Latin Square \cite{richardson2018use}, with four participants assigned to each combination randomly. %Navigation time was measured starting with participants’ first step and ending when they arrived within one meter of the destinations.

After each navigation, participants described the route with landmarks they recalled. They also evaluated the system performance using UMUX-LITE \cite{lewis2013umux} on \textit{Perceived Effectiveness} and \textit{Ease of Use}, and NASA Task Load Index (TLX) \cite{hart1988development} on \textit{Mental Demand}, \textit{Physical Demand}, \textit{Performance}, and \textit{Frustration}. Participants gave a 7-point Likert scale on each subjective measure. See Supplementary Section 7
%\ref{supp-sec:evaluation-questionnaire-interview-questions} 
for the full script.
%including \textit{Perceived Effectiveness} (\ie{} This system's capabilities meet my requirements for navigation), \textit{Ease of Use} (\ie{} This system is easy to use for navigation), \textit{Mental Demand} (\ie{} The navigation task was very mentally demanding), \textit{Physical Demand} (\ie{} The navigation task was very physically demanding), \textit{Self-rated Performance} (\ie{} I accomplished the navigation task I was asked to do very successfully), and \textit{Frustration} (\ie{} I felt very frustrated during this navigation task). Participants rated each measure using a 7-point Likert scale, with 1 being totally disagree and 7 being totally agree.
We also collected participants’ qualitative feedback on both systems.

\subsubsection{Free-Form Exploration}
\colorchange{Participants then conducted a free exploration task within the square using \system{}. They were contextualized in the scenario where they have some time to explore the square and existing annotations at their own pace before their friends arrived, and were encouraged to create new annotations for audiences of their choosing. During the exploration, they could freely use all features of \system{} to navigate and query their surroundings. Each participant was required to create at least three annotations.}

\subsubsection{Exit Interview}
We ended our study with a semi-structured interview, discussing annotations participants created, the purposes and target audiences of these annotations, as well as annotations and scenarios they envisioned outside the square (see Supplementary Section 7%\ref{supp-sec:evaluation-questionnaire-interview-questions})
. We also asked for feedback on existing annotations, on the design of \system{}, and suggestions for future \insitu annotation authoring systems. %See the Supplementary Materials for all questionnaires and interview questions.

\subsection{Analysis}
\label{subsec:analysis}
%We collected both quantitative measures and qualitative feedback from the study. 
Since two participants (P17, P18) did not complete the navigation task (Section \ref{subsec:evaluation_participants}), we analyzed the quantitative data from P1–P16 while including qualitative data from all 18 participants. \colorchange{For P1–P16, the navigation task routes and the system order were fully counterbalanced, with four participants for each combination of the route and route order.}
Our analysis methods are described below.

\subsubsection{Measures \& Statistical Analysis}
In addition to the Likert scale ratings (Section \ref{subsec:procedure}), we defined two measures for navigation: (1) \textit{Navigation Success}, a binary value indicating whether a participant successfully reached the destination 
and (2) \textit{Total Landmarks Recalled}, the number of landmarks participants correctly remembered, reflecting their mental map construction \cite{chen2025visimark}. In line with established definition that landmarks should be permanent, unique and identifiable \cite{yesiltepe2021landmarks}, we considered flower bed, statue, stone podium, and table tennis table as landmarks, while excluding transient or repetitive features such as people, benches (20 in the square), and trash bins (26 in the square).

Using the Shapiro-Wilk test, we found none of the measures to be normally distributed. Therefore, we applied a generalized linear mixed model (GLMM) \cite{bolker2009generalized} to \textit{Navigation Success} and Aligned Rank Transform (ART) ANOVA on other measures for statistical analysis.
For each measure, we included one within-subjects factor, \textit{System}, with two levels: \system{} and \baseline{}. 
To validate the counterbalancing, we also included \textit{Order} representing the order of the two systems in the navigation task and found no significant effect of \textit{Order} on any measures.  \colorchange{We evaluate statistical significance at $\alpha = .05$.} Partial eta-squared ($\eta_p^2$) was used to calculate effect size, with 0.01, 0.06, 0.14 representing the thresholds of small, medium and large effects \cite{cohen2013statistical,wang2024gazeprompt}.

\subsubsection{Qualitative Analysis}
The thematic analysis process \cite{braun2006using,clarke2017thematic} for this study was identical to that in the Formative Study (Section \ref{subsec:formative_analysis}), except that the initial open-coding by two researchers was performed over five participants' transcripts (\(\sim \)28\% of the data).
In addition to participants' transcripts, the research team analyzed all user queries to categorize user questions. We only include questions initiated by participants and exclude those prompted by the research team during the tutorial (Section \ref{subsec:procedure}).
We also analyzed the system responses of all user queries regarding response correctness and types of error.

\section{Evaluation Study Results}\label{sec:evaluation_results}
We report participants' performance in the navigation task and experiences with \system{} across navigation, exploration, and user query. We then share participants' envisioned future system use cases.

\subsection{Navigation Experiences}
\subsubsection{\system{} Improved Navigation Performance}
We found a significant effect of \textit{System} on \textit{Navigation Success} (\colorchange{$\chi^2(1) = 9.11$, $p=0.003$}), with 14 of 16 participants successfully navigating using \system{} (P1-P9, P11-P14, P16) compared to only 6 with the baseline, \baseline{} (P4, P5, P7, P13, P15, P16). Of the two participants (P10, P15) who did not complete navigation with \system{}, P10 stopped about two meters from the destination due to phone tracking drift (Section \ref{subsubsec:Detecting and Notifying Users of Drift}), and P15 considered a similar nearby object as the destination and stopped following instructions.

%TC:ignore
\begin{figure*}[htb]
    \centering
    \begin{subfigure}{0.6\textwidth}
        \centering
        \includegraphics[width=\textwidth]{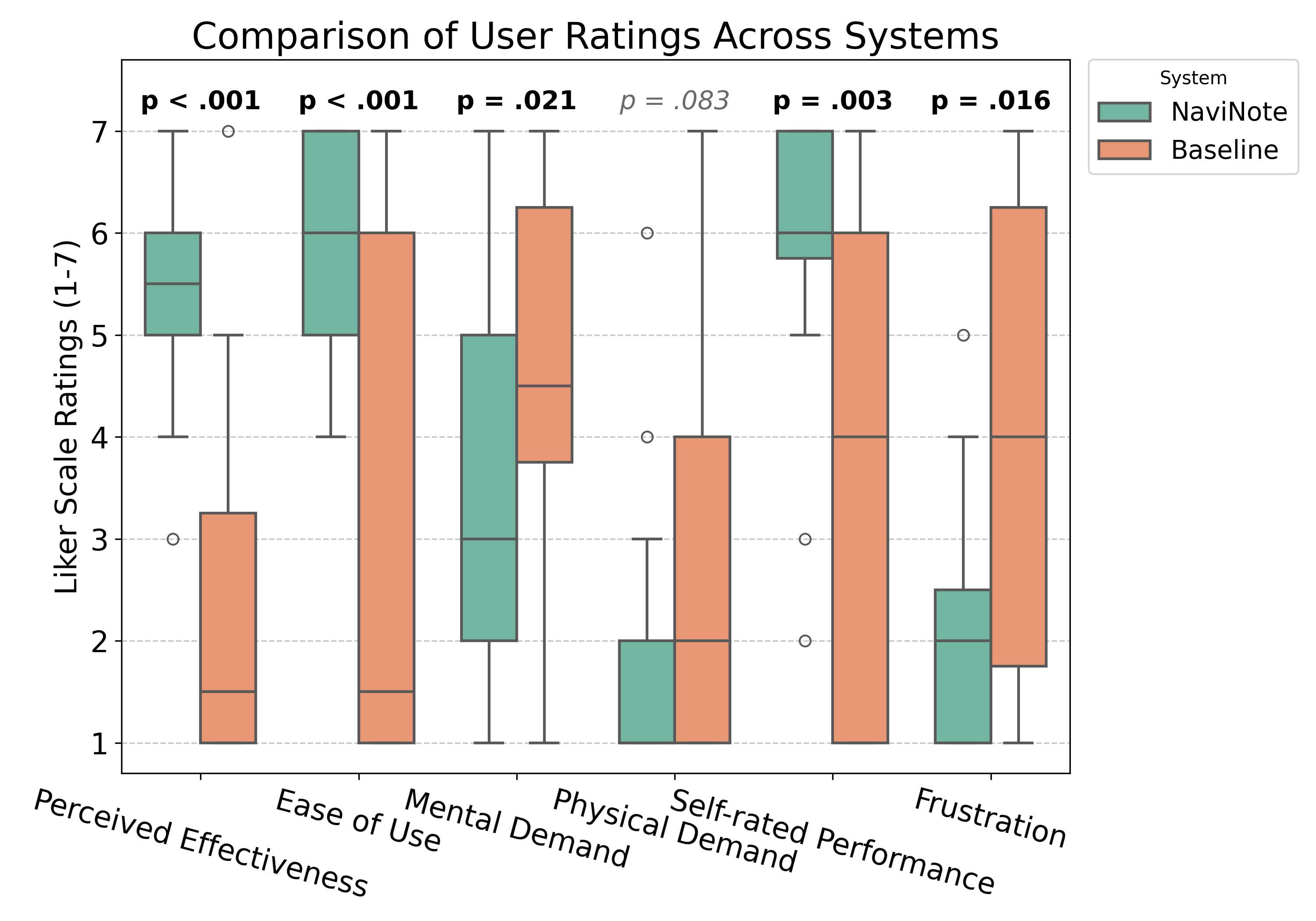}
    \end{subfigure}
    \caption{Comparison of users’ subjective ratings between NaviNote and the baseline system using UMUX-LITE (Perceived Effectiveness, Ease of Use) and NASA-TLX (Mental Demand, Physical Demand, Self-rated Performance, Frustration)}
    \Description{In the box-plots provided, all subjective ratings are significantly superior for NaviNote compared to the baseline, except Physical Demand, where there is no significant difference. For Physical Demand, the baseline ratings' box and whiskers span the entire Likert Scale, with the median being 2, a low physical demand; whereas the NaviNote ratings' box and whiskers span from 1 to 3, with two outliers at 4 and 6. NaviNote's median is at 1, the lowest physical demand score possible.}
    \label{fig:likert_scale}
\end{figure*}
%TC:endignore

\subsubsection{Easier Navigation Perceived with \system{}}
We found a significant effect of \textit{System} on \textit{Mental Demand} \colorchange{($F_{1,15}=5.36$, $p=0.035$, $\eta_p^2=0.26$)}, \textit{Self-rated Performance} \colorchange{($F_{1,15}=15.71$, $p=0.001$, $\eta_p^2=0.51$)} and \textit{Frustration} \colorchange{($F_{1,15}=12.00$, $p=0.003$, $\eta_p^2=0.44$)}, as shown in Figure \ref{fig:likert_scale}. Compared to navigating using \baseline{}, participants felt that navigating with \system{} was less mentally demanding \colorchange{($diff=-6.12$, $p=0.035$)}, more successful \colorchange{($diff=8.69$, $p=0.001$)}, and less frustrating \colorchange{($diff=-7.38$, $p=0.004$)}. %, signaling they considered navigation easier with \system{} than \baseline{}.
Fifteen participants appreciated \system{}’s continuous instructions. 
Eight participants considered the instructions effective for reducing mental load, as the constant confirmation provided ``a lot of support and comfort'' (P14). In contrast, fourteen participants felt that \baseline{} was insufficient for navigation, noting that they needed to take ``indefinite number of photos'' (P10, P14) and only received directions by chance (P6, P10, P14).

While no significant effect was found in \textit{Physical Demand} (\colorchange{$F_{1,15}=3.14$, $p=0.097$}, $Mean_{\text{\system{}}}=1.82$, $Mean_{\text{\baseline{}}}=2.59$, $SD_{\text{\system{}}}=1.42$, $SD_{\text{\baseline{}}}=2.00$), four participants (P8, P10, P11, P15) reported that taking photos while walking increased their physical demand. P5 and P13 further explained that with \baseline{}, the lack of directional information led them to take smaller steps, which in turn increased the physical load. Nine participants appreciated \system{} for being hands-free, and eight liked \system{} for no need to aim the camera at specific targets. As P13 detailed, ``So with TapTapSee, it's expecting you to be able to point the camera at something and being able to find what you're looking at. With [\system{}], I don't have to look.'' Compared to \baseline{}, which only had access to the camera frame, \system{} incorporated the spatial understanding of the entire POI beyond camera view, an essential part of making \system{} hands-free.
Additionally, nine participants disliked \baseline{} for having to stop and perform additional gestures while walking with white canes, describing this process as ``completely disorientating'' (P7).

\subsubsection{Design Considerations for Navigation Cues}
We found a significant effect of \textit{System} on \textit{Perceived Effectiveness} \colorchange{($F_{1,15}=38.44$, $p<0.001$, $\eta_p^2=0.72$)} and \textit{Ease of Use} \colorchange{($F_{1,15}=26.00$, $p<0.001$, $\eta_p^2=0.63$)}, where participants considered \system{} more effective for navigation \colorchange{($diff=12.9$, $p<0.001$)} and easier to use \colorchange{($diff=10.2$, $p<0.001$)} than \baseline{}.
Participants shared their experience of the audio compass, verbal instructions, and multimodal feedback on safety warnings. They also offered improvement suggestions:

\textbf{\textit{Audio Compass.}} Eight participants specifically mentioned appreciating the audio compass, describing it as easy to understand (P10, P13) and helpful for finding and confirming the correct direction (6/18). Four participants (P7, P14, P17, P18) further suggested that, in addition to louder beeping signals when facing the correct direction, \system{} should also provide real-time verbal confirmation.
In terms of distraction, P15 found the compass subtle and not distracting, whereas P2 and P16 found it confusing and preferred only verbal corrections.

\textbf{\textit{Verbal Instructions.}} Eight participants found the turn-by-turn instructions straightforward and easy to follow. Participants exhibited different preferences on the information within the instructions: four (P6, P8, P11, P14) appreciated constant directional assurance, and five liked the occasional distance-to-destination updates, which assured them that they were ``getting closer'' (P7). However, three participants (P9, P13, P18) found the updates on distance distracting and wanted only immediate direction.
P9 and P18 liked the instructions for providing information about ``haptic path'' (Section \ref{subsubsec:navigation}). As P9 explained, ``If I go anywhere, I try to check if there is a wall or a curve or something to follow to keep in straight line. [... In the experiment I] followed the flower bed, which was a good guideline.''

Four participants (P7–P9, P12) noted a learning curve for the instructions. Five participants wanted \system{} to explicitly say when they were going the wrong way, rather than just telling them to ``turn backward.'' They also suggested using more natural phrases, such as ``turn around'' instead of ``backward'' (5/18), and simplifying egocentric directions to four basic options (\ie left, right, forward, backward) rather than eight (P8, P9).

\textbf{\textit{Multimodal Feedback.}} \system{} provided haptic vibrations when playing safety annotations (Section \ref{subsubsec:annotation_authoring}). P13 appreciated the multimodal feedback as it made the system more robust: ``I wasn't just relying on one thing. If something stopped working but I felt a vibration, I still knew there was something happening or there was something to be aware of.''

\subsection{Annotation Experiences}

\subsubsection{\system{} Helps Users Understand their Surroundings.}
\label{subsubsec:help_understand}
We found a significant effect of \textit{System} on \textit{Total Landmarks Recalled} (\colorchange{$F_{1,15} = 15.21$, $p = 0.001$, $\eta _p^2 =0.50$}), with participants remembering more landmarks with \system{} than \baseline{} (\colorchange{$diff=8.75$, $p=0.001$}).
Thirteen participants mentioned how the annotations in \system{} helped them learn about their surroundings, \eg by providing general descriptions of the square and by offering specific information tailored to their needs. Nine participants praised the annotations for giving a general impression of the layout and available amenities. As P15 recounted, ``[with \system{}] I learned there are branches, there are benches, there are statues, there are flowers, and people sitting around.'' P10 further noted that crowdsourced annotations in \system{} could help identify ``what services or amenities are available'' and could help him plan for social activities.

In terms of specific information, four participants (P1, P4, P7, P14) noted appreciating historical annotations about the square and the statue. Three participants (P5, P13, P18) valued information about local attractions, such as descriptions of flowers and their species (P13, P18), which raised their awareness of features they would otherwise have difficulty seeing (P5, P9). 
Notably, four participants (P12, P15, P17, P18) praised \system{} for providing color information about objects. P12 noted that color information about obstacles with low contrast against the background (\eg a gray raised flower bed \vs{} gray ground) made them easier to detect and avoid.

\subsubsection{Experiences with Different Types of Annotations}
\label{subsubsec:experience_annotations}

In addition to agreeing that the layout and amenities annotations helped them understand their surroundings (Section \ref{subsubsec:help_understand}), participants highlighted the help of safety and accessibility annotations in supporting safer mobility. Eleven participants noted appreciating these annotations during navigation and free exploration, as they provided advanced alerts of obstacles, helping them avoid them. As P14 described, ``[with safety annotations] it was good to know even before I get there that I'm going to encounter a raised [flower] bed.'' Three participants (P3, P5, P10) noted liking \system{} for \textit{automatically} playing safety alerts. Four participants (P1, P7, P10, P17) further emphasized the usefulness of warnings about overhanging obstacles (\eg{} tree branches), which conventional aids, like white canes, typically miss. As P1 explained, ``[\system{}] would be a complementary assistance to white cane... because the idea is to avoid bumping my head into a barrier.''

Participants expressed diverse opinions on attraction and experience annotations. Several participants (6/18) appreciated attraction annotations for sparking interest in nearby areas rather than ``easily missing them'' (P18), and three (P4, P8, P17) even considered them their favorite type of annotation. In contrast, four participants (P3, P5, P6, P10) chose to filter out attraction annotations due to lack of interest or usefulness. Notably, all four of these participants were blind. Opinions on experience annotations were similarly mixed: three participants (P1, P8, P16) liked how they shared personal experiences of the area, especially from BLV users' perspectives (P8), while P10 preferred practical information over subjective experiences, and P5 noted only being interested in such annotations from friends.

\subsubsection{Annotation Authoring}
\label{subsubsec:results_annotation_authoring}
All participants created annotations successfully. \colorchange{Building on existing annotations and querying \system{} for additional information, they actively created new annotations to share with themselves as reminders (4/18), with friends and family (7/18), with the BLV community (8/18), and with the public (7/18).} In addition, participants (12/18) appreciated the ability to edit and delete annotations, as this allowed them to correct mistakes (P8, P12, P17) and prevent AI misunderstandings (P1, P10). Most annotations that participants created fit within our existing annotation taxonomy: safety (6/18), accessibility (P7), amenity (7/18), layout (5/18), attraction (P2, P8, P9), and experience (8/18). Interestingly, three participants (P4, P13, P14) created annotations that aimed to \textit{request information from other users}. For instance, when feeling unsure about the identity of a statue, P14 created an annotation asking other users to ``find it out'' and tell her. This type of annotation fell outside of our original categorization, but would nonetheless provide valuable opportunities for users to communicate with each other; thus, we add a final category to our taxonomy, ``Request.''

\subsubsection{Control Over Annotation Access}
Six participants appreciated that \system{} allowed them to filter annotations by type. Seven participants suggested refinements to these %\textbf{
``information layers'' with respect to annotation category, content, and context. P7 proposed finer-grained annotation categorization, while both P3 and P7 recommended filtering by content rather than type to retrieve useful information. For non-safety annotations, P10 wanted annotations linked to places of personal significance to automatically play, P14 suggested applying location-aware filtering (\eg outside the square \vs in the square), P13 and P16 recommended merging duplicate information (\eg flowers in two identical flower beds), and P11 suggested accessing these annotations only in first-time visits.

While participants valued the different access modes for annotations (\ie automatic, prompted, or silent), P13 found asking the system for the prompted annotations to be tedious. 
She suggested incorporating lightweight gestures for annotation access, such as using the volume button to quickly play or dismiss prompted annotations.

\subsection{User Query Experiences}
\label{subsec:user_query_experiences}
\colorchange{We evaluate the performance of \system{}'s voice-based question answering regarding the accuracy of \system{} and participants' interaction experiences. We further analyze the types of user query and \system{}'s response time}.
% All quotes are taken from the experiment logs.

\subsubsection{Accuracy.} Participants asked a total of 605 questions. Among these questions, 501 (82.8\%) were correctly answered, 34 (5.6\%) were incorrect, 13 (2.2\%) were partially correct, 39 (6.5\%) were misrecognized by the speech-to-text model (\eg due to accents, background noise or low volume), and the AI failed to answer the remaining 18 questions (3.0\%) for various reasons listed below.

Half of the incorrect answers (17/34, 50\%) were caused by failing to retrieve relevant information for generating answers or invoking actions. These failures stemmed from miscommunication between the orchestrator and the agents, such as not passing or returning the correct parameters (\eg when a user asked about the statue’s history, the orchestrator only instructed the annotation agent to search for history-related annotations without specifying statue-related ones). The second most common cause was the orchestrator invoking the wrong agent as it misunderstood the user query due to two similar terms (\eg ``attractions'' \vs ``attraction annotations,'' which prompted the orchestrator to invoke the internet search agent or annotation agent respectively; 6/34, 17.6\%). Other reasons included the AI failing to retrieve chat memory for context-based answers (5/34, 14.7\%), retrieving incorrect information (\eg using users' Unity coordinates instead of real-world positions; 5/34, 14.7\%), and AI hallucinations (1/34, 3.0\%)
For the 13 partially correct answers, the most frequent cause was the orchestrator generating the correct answer without invoking the appropriate agent to trigger the corresponding frontend action (\eg start navigation; 8/13, 61.5\%). 

For the 18 questions that \system{} failed to answer, eleven (61.1\%) occurred because the system lacked available information (\eg the participant asked ``How is the weather today?'' while \system{} has no access to the date). Six cases (33.3\%) were due to ``Too Many Requests'' errors from OpenAI API, and one case was due to OpenAI incorrectly filtering safe content.

\subsubsection{Conversational Voice-based Interaction Experiences.} Thirteen participants explicitly appreciated \system{} for the conversational and voice-based interaction design, describing it as ``intuitive and user-friendly'' (P9). P7 also said, ``I've never seen a system where you can say instructions. You [usually] have to go into menus to do things.'' P17 further suggested that, in addition to voice control, the system should incorporate a graphical user interface to allow exploring available functions with screen readers. 

We also observed that 11 out of the 18 participants interacted with the system in a conversational style. For example, when creating an annotation, instead of immediately providing the content, participants would first tell the system they wanted to create a annotation, wait for the system to ask for content, then provide it in a follow-up query. Five participants also asked \system{} to repeat responses.

\subsubsection{Question Types.}
As participants often asked similar questions repeatedly, we counted the number of participants who asked each type of question rather than the total occurrences, to prevent any single participant from disproportionately affecting the distribution of question types.
Among all questions initiated by the participants, the most common three types were directly related to system features: (1) All 18 participants inquired about annotation content and asked to author annotations; (2) Seventeen participants asked to customize \system{}; and (3) Twelve participants asked about navigating to specific locations. 
In addition to questions related to system features, participants asked four types of questions related to their surroundings. Ten participants (55.6\%) asked for \textbf{general descriptions of their surroundings}, such as the layout and size of the square.
% (\eg ``Can you tell me how large the square I am in is?'').
Nine participants inquired about the \textbf{objects and activities at specific locations}, such as number of benches and available services at the square.
% (\eg ``How many benches are there in the square?'', ``What services and activities [are] available in the park?'').
Nine participants asked about \textbf{object details} in or beyond the camera view. For example, P17 asked, ``Could you describe these birds for me?'' after hearing bird sounds.
Eight participants asked about \textbf{object locations relative to themselves} (\eg 
% ``What food and dining places are in close proximity to my location?'',
``Is the flower bed [that] the one attraction note [talks about] on my left or right?'').

\subsubsection{Response Time.} 
\begin{figure*}[htb]
    \centering
    \begin{subfigure}{0.99\textwidth}
        \centering
        \includegraphics[width=\textwidth]{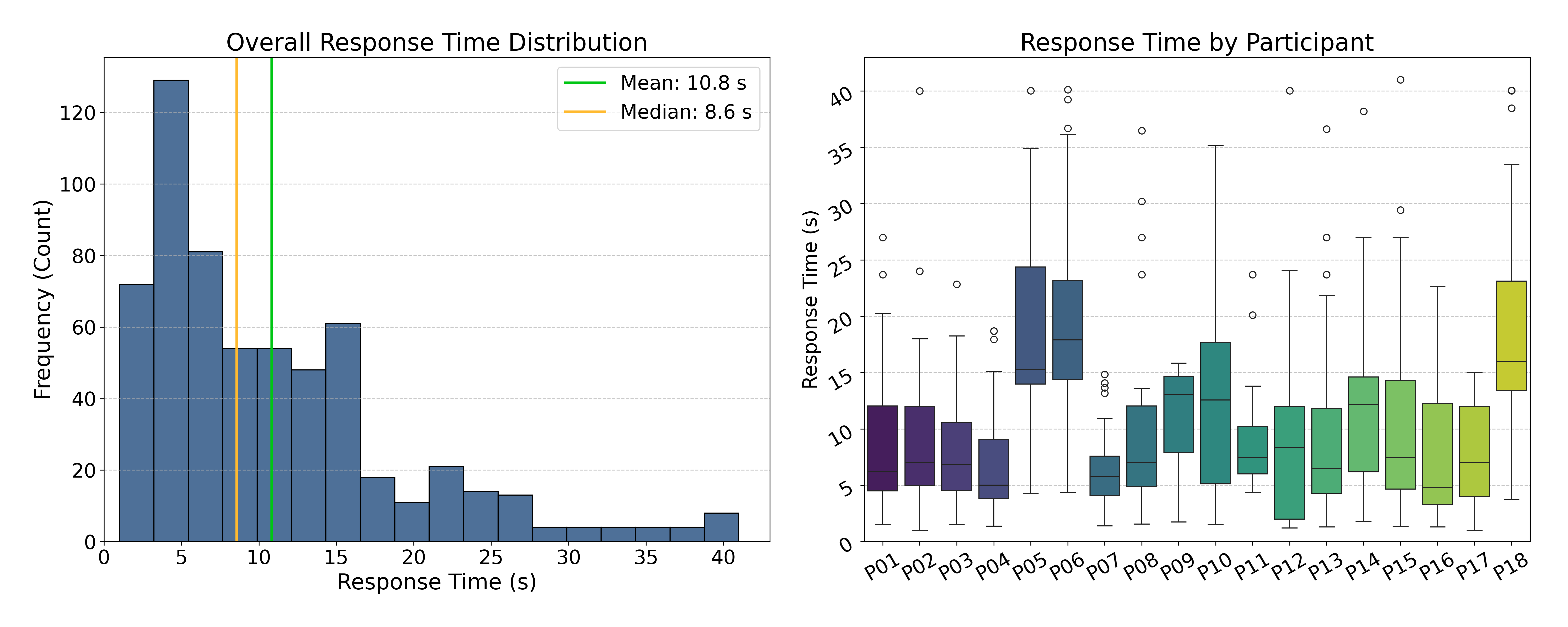}
    \end{subfigure}
    \caption{Left: Distribution of \system{}’s response time per user query, with mean response time to be 10.8 seconds and median response time 8.6 seconds. Most queries were answered within 15 seconds. Right: Response times for individual participants, with variation between participants primarily due to unstable internet connections in the field. } 
    \Description{Left: Histogram of \system{}’s response time per user query. The most frequent values clustered around 5 seconds. Right: Box-plot of system response times for individual participants. Most participants’ times ranged between 3 and 15 seconds. P5, P6, and P18 showed longer response times, around 15–25 seconds.}
    \label{fig:response_time}
\end{figure*}

\system{} answered user queries in an average of 10.8 seconds ($SD=8.3$, $Max = 41.0$, $Min = 1.2$, see Figure \ref{fig:response_time}). The variation was partly due to question complexity (\eg ``Hello!'' \vs ``My friend Emma left me a note. Can you help me find where she is?''), %\ggh{Is this example a supported capability of the system?},
while longer delays were typically caused by unstable internet or occasional ``Too Many Requests'' errors from the OpenAI API, upon which the API waited for around 20 seconds before retrying. Six participants (P2, P5, P6, P9, P17, P18) expressed dissatisfaction (\colorchange{\textit{Mean response time }$=15.1 s$}), noting that the system should respond more quickly, while P15 (\colorchange{\textit{Mean response time }$=10.8 s$}) was satisfied and described it as ``responding very quickly.'' 
%TC:ignore
%TC:endignore

\subsection{Independence through Annotations: When Are Annotations Useful?}
Eight participants highlighted that \system{} gave them \textbf{more independence} compared to asking others for help. As P13 explained, ``I don't want to always have to wait for somebody sighted. I quite like to get out, but people always get worried.'' Participants appreciated that \system{} enabled them to take initiative rather than receive information passively (5/18), by asking follow-up questions and actively leaving notes. Most participants (14/18) noted that they would feel comfortable exploring alone with \system{} in the future.

Participants also envisioned potential use cases for \system{} beyond the experiment. These cases generally fell into four categories: navigation, relaxation and exploration, information gathering, and social interactions.

\subsubsection{Navigation.} Several participants (8/18) envisioned using annotations to mark interested spots and navigate back in later days. For example, five participants wanted to mark indoor public spaces, such as airport gates, train stations, and malls for easier access. P7 also proposed returning home with \system{}: ``My biggest fear is getting stuck in the house, not able to come out because I've always got this fear: The minute I leave my house, I won't be able to find my house again because all the houses look the same and feel the same from outside.''

\subsubsection{Relaxation and Exploration.} Participants wanted to use \system{} for museum exploration, street wandering, and travel (8/18) to help independent exploration, which is often challenging. For example, P5 said, ``Usually I would just walk [past exhibits]. That's why I hate going to museums. There's a load of stuff I can't see.'' With \system{}, however, she could discover visual attractions that she normally wouldn't see.

\subsubsection{Information Gathering.} Participants saw potential value in viewing others' comments in advance. As P4 envisioned, ``I can see a map of [annotations] and then there's like little stars or something. I click it and I can hear it. I would actually do that.'' They also wanted to gather information about tourist attractions (P5), or leave notes to request information, so that other BLV individuals visiting the same location later could benefit (P13, P14).

\subsubsection{Social Interactions.} Participants envisioned leaving audio notes for social purposes (6/18), such as marking meeting locations or sharing information with friends. For example, P13 imagined a scenario where, if she were running late, her friends could leave her an annotation about a new meeting spot. P16 described annotations as a feature that ``makes [the environment] more interactive'' by sharing information with other users, and P4 further emphasized that accessible annotations can promote social interaction: ``Can you imagine that? You can actually interact with people. Because a lot of people, they don't interact that much... So I like the audio note element [for allowing more interactions].''

\section{Discussion}
Our evaluation showed that \system{} effectively supports BLV participants in \colorchange{independently authoring annotations}, navigating the last few meters to POIs, and exploring local public places. Participants also envisioned using it across various use-cases. In this section, we discuss design challenges we discovered related to spatial annotation authoring tools, reflect on agent-based accessibility systems, and outline limitations and future directions. Note that in this section we discuss broad topics, while we outline specific design implications for \system{} in Section \ref{sec:evaluation_results}.

\subsection{Design Challenges for Spatial Annotation Authoring Tools}
Participants praised \system{} for supporting scene exploration and enabling them to author and access spatial annotations. They also suggested ways to improve retrieval mechanisms and safety, as discussed below. 

\subsubsection{How can annotations maintain scene alignment over time?} \label{subsubsec:How to keep annotations aligned with scenes?}
While crowdsourced annotations anchored to spatial locations can help BLV users explore their surroundings, they also introduce the challenge of staying physically aligned after changes in the scene. During free exploration, participants were concerned that objects of interest may be moved or removed, making associated annotations obsolete. In addition to manual edits via crowdsoucing, advances in 3D change detection \cite{Mandal20193DFR,Mandal20203DCD, Qiu20233D} opens up the possibility of automatic detection and realignment of annotations \cite{numan2025adjustar}. We suggest that future systems explore automatically detecting environmental changes and realigning or removing outdated annotations to ensure they remain reliable.

\subsubsection{\colorchange{How can annotations be kept true, safe, and ethical?}} \label{subsubsec:discussion_ethical_annotations}
As with many user-generated content (UGC) platforms, crowdsourcing introduces risks of rumors and malicious content, which can undermine users' experiences and even threaten safety \cite{cheng2015antisocial,sharevski2023designing, sharevski2023just}, especially for BLV users. Existing mechanisms from UGC platforms can be adapted to moderate such annotations: (1) ``Official/Verified'' annotations from trusted users, similar to the ``blue checkmark'' system on X \cite{X:verified} or Instagram \cite{Instagram:verified}; (2) Upvote/downvote mechanisms, where heavily downvoted annotations are removed \cite{lampe2006ratings}; and (3) AI filtering to automatically detect and remove unsafe notes. That said, AI-based approaches are not always reliable due to inaccuracies and biases in training data \cite{binns2017like, anigboro2024identity}. For instance, in our study, AI incorrectly filtered out a harmless note about a nearby mosque (1/605). Future work should explore strategies to ensure spatial annotations remain genuine, safe, and ethical.

\subsubsection{How to manage annotation access and prevent information overload?}
\label{subsubsec:discussion_privacy}
\colorchange{In the current prototype of \system{}, all notes were made publicly accessible to simplify evaluation. However, our findings show that participants frequently authored annotations intended only for themselves, their acquaintances, or specific BLV audiences (Section~\ref{subsubsec:results_annotation_authoring}), which further supports the social annotation needs reported in FootNotes \cite{gleason2018footnotes}. To better protect end-user privacy, future systems should incorporate fine-grained access control mechanisms that verify the intended audience before an annotation is shared. Moreover, systems should offer options for anonymity and warn users when they disclose sensitive or identifiable information \cite{krol2016control}. For example, an annotation such as ``Meet me here'' should only be visible to trusted contacts, and a note like ``This is my favorite cafe in my neighborhood'' should be shareable anonymously.}

\colorchange{In addition to privacy considerations, filtering information based on target groups can help prevent information overload. \system{} currently prioritizes safety- and accessibility-related annotations and skips notes that cannot be played within 15 seconds after a user approaches them (Section~\ref{subsubsec:annotation_authoring}). Future systems should further prioritize annotations based on both the identity of the contributor (\eg assigning higher priority to notes from known or trusted people) and community ratings within the target group (\eg upvotes or verification mechanisms as discussed in Section~\ref{subsubsec:discussion_ethical_annotations}). These strategies can help ensure that users receive relevant, trustworthy, and manageable information.}

\subsection{Usability Challenges of Exploration Systems in the Wild}
\label{subsec:wild_usability}
While \system{} provided substantial support in last-few-meters navigation and free exploration, our evaluation also revealed several usability challenges, which we reflect on below.

\subsubsection{Crowdsourced Scanning}
\system{} is built on the assumption that scans can be crowdsourced from sighted volunteers and friends, a process that has been used to capture locations world-wide \cite{nummenmaa2025employing, niantic2025wayfarer, brachmann2025buildingspatialintelligence}, and can be completed online \cite{niantic2025vpsactivated} and offline \cite{niantic2025devicemaps}. With consent, BLV user sessions could also be used to keep scans up-to-date. In the future, we also envision gathering scans from drones \cite{ngo2023photogrammetry,yu2024automated}, delivery robots \cite{matthies2002portable}, and automated vehicles \cite{belkouri2022through}. Given VPS scans still have limited coverage \cite{brachmann2025buildingspatialintelligence}, in the meantime, research should investigate methods to provide seamless transitions between more/less accurate localization methods, \eg between VPS and GPS, such that users can have \textit{at least some} guidance no matter where they are. Furthermore, research should explore mechanisms to automatically update scene understanding, \eg whenever users localize and contribute new scans (Section \ref{subsubsec:How to keep annotations aligned with scenes?}).

\subsubsection{Guidance for Independent Localization}
Despite not needing to aim \system{} at specific targets during use, \system{}'s VPS still requires an initial localization, in which users sweep the camera back and forth until the system identifies its location \cite{lightshipVisualPositioning,google2025geospatial}.
Although we did not formally test this localization process with every user, as it was out of scope for our study, ten participants tested independent localization with eight succeeding within one minute (80\%). Future work should explore how to improve localization for BLV users \colorchange{to enhance the generalizability and robustness of our findings}, \eg by designing effective audio and haptic guidance for this process similar to \cite{chung2021understanding, dontlooknow, menelas2014non, rahman2023take}.

\subsubsection{Detecting and Notifying Users of Drift}
\label{subsubsec:Detecting and Notifying Users of Drift}
While VPS enabled high accuracy and continuous localization, some participants still experienced mid-use drift, where the phone failed to track its 3D position accurately, causing the navigation algorithm to misguide them \colorchange{(\eg P10 stopped about two meters from the destination)}. \colorchange{To accurately reflect \system{}’s performance, we did not intervene when drift occurred.} Even though participants were instructed before the task to navigate as they normally do (\eg with white canes) and only treat \system{}'s instructions as an additional reference point, they still exhibited a high-level of trust in the system during the experiment. 
Future tools should explore mechanisms to counter mid-use drift (\eg{} visual-based pose estimation \cite{Li2017Monocular} and localization \cite{Ai2020Visual}, reprojection \cite{dontlooknow}), as well as using MLLMs for visual-semantic reasoning to correct AR content \cite{numan2025adjustar}. Moreover, future research should explore effective ways to detect and alert users of drift, such as reporting localization confidence and instructing users to re-localize. \colorchange{Annotating the environment with landmark-specific cues (\eg haptic textures, smells, or characteristic sounds \cite{saha2019closing, loomis2001navigating}) may also improve landmark recognition and further support users’ self-localization when drift occurs.}

\subsection{AI Agent-based Systems for Accessibility}
\label{subsec:agent_for_a11y}
Most of our BLV participants expressed trust in \system{}. As P8 noted, ``The way it gives directions is very clear. And then that's how it builds up trust.'' However, there were \colorchange{one case (1/605)} where the system generated incorrect answers and the participant accepted fabricated information (although this was rare). For example, P2 believed the statue depicted someone other than who it was based on a system hallucination. Future systems should make users aware of AI's limitations (\eg the types of questions it can reliably answer) and the possibility of hallucinated responses before use \cite{kaate2025you,hong2024understanding,nahar2024fakes,nahar2025catch}. They should also investigate how to provide richer context and details to help BLV users recognize errors independently \cite{hong2024understanding}, such as including information sources and image descriptions in the response %to help users verify a statue’s identity 
(\eg ``This information came from the internet'' \vs ``... from user X's annotation''). More importantly, future work should focus on building more reliable and explainable AI systems \cite{dalrymple2024towards,arrieta2020explainable,dwivedi2023explainable}.

AI systems can also suffer from response latency \cite{lee2025imaginatear, zhang2024explainingthewait}. In our study, some participants expressed wanting faster responses (Section \ref{subsec:user_query_experiences}). Since \system{} incorporated the ``Orchestrator'' pattern \cite{openai_agent}, response time was influenced by both the base model latency \cite{achiam2023gpt} and question complexity, which could require one or multiple rounds of communication between the ``Orchestrator'' and specialized agents. Beyond model-level improvements, future research should also investigate speed optimization of agent-based architectures, while tailoring responses to BLV users’ needs \cite{gulli2025agents}.

\subsection{Limitations and Future Directions}
This work has several limitations and future directions, in addition to those mentioned in the discussion above. First, although the pre-set annotations used in our system evaluation were derived from participants’ desired annotations identified in our formative study, they were tested in single-user sessions. As a result, our evaluation may not fully capture the dynamics of community-based, crowdsourced annotations, \eg how BLV users might interpret colloquial or incomplete annotations. Moreover, we evaluated \system{} in a local public square with comparatively few objects and relatively simple routes, which may not fully reflect BLV users' actual daily settings.
Additionally, further effects and concerns may have eluded us based on our relatively small sample size of 16 users.
In summary, future research should investigate BLV users' experiences with crowdsourced annotations in longitudinal and multi-user studies, test such systems in more complex and realistic public environments, and include more participants with diverse visual conditions.
\section{Conclusion}
Thanks to a formative study with 24 blind and low vision (BLV) participants, we contribute the design and implementation of \system{}. \system{} is an accessible voice-based interface, incorporating Visual Positioning Systems and Multimodal Large Language Models to enable BLV users to author \insitu annotations and navigate effectively in last-few-meters scenarios. Through an evaluation with 18 BLV participants, we found that \system{} significantly improved BLV users’ navigation performance, helped them better understand their surroundings, and enabled them to actively annotate the environment. We also discuss design considerations for future accessible annotation systems and provide an initial taxonomy of BLV users’ desired spatial annotations to guide future crowdsourcing efforts.

\begin{acks}
We thank Bryan Tan, Sara Gomes Vicente, Michael Firman, Jakub Powierza, Chris Shaw, Michael Miller, Ben Benfold, Alyssa Coley, Effie Shum, and Belinda Manfo from Niantic Spatial, Inc. for their assistance in the successful execution of both studies and in system deployment. We thank \href{https://beyondsightloss.org.uk/}{Beyond Sightloss} for facilitating participant recruitment in both studies.
\end{acks}

\bibliographystyle{ACM-Reference-Format}
\bibliography{sections/references}

\newpage
\appendix
%TC:ignore

\section{Participant Demographics in the Formative Study} \label{sec:formative_demographics}
\begin{table*}[htbp]
\scriptsize
\centering
\caption{Participant demographics in the formative study.}
\label{tab:demographics_formative}
\begin{tabular}{>{\raggedright\arraybackslash}m{0.4cm}>{\raggedright\arraybackslash}m{1.6cm}>{\raggedright\arraybackslash}m{1.0cm}>{\raggedright\arraybackslash}m{2cm}>{\raggedright\arraybackslash}m{2.9cm}>{\raggedright\arraybackslash}m{2.8cm}>{\raggedright\arraybackslash}m{2.3cm}
% >{\raggedright\arraybackslash}p{2.5cm}
}
\Xhline{2\arrayrulewidth}
  \textbf{ID} & \textbf{Day and Group} &\textbf{Age/Gender} &  \colorchange{\textbf{Visual Condition}}& \textbf{"Can you match clothes/objects according to their color?"}&\textbf{"Can you notice objects around you as you walk?"}&\colorchange{\textbf{If you were paired with a totally blind participant, would you be able to help guide them?}}\\
\Xhline{2\arrayrulewidth}
F1 & Day 1, Group A&45/F & Low vision&Sometimes&When close by&N\\
\hline
F2 & Day 1, Group A&43/M & Low vision&TapTapSee tells me colors of things&Yes with good contrast&Sometimes\\
\hline
F3 & Day 1, Group A&20/F & Low vision &N&If it’s far then no&Y \\
\hline
F4 & Day 1, Group A&33/F & Low vision, macular dystrophy &Y&If it’s near yes but if it’s far it harder to see&Y\\
\hline
F5 & Day 1, Group B&58/F & Totally blind&N&N&N\\
\hline
F6 & Day 1, Group B&25/F & Low vision&Y&Y&Y\\
\hline
F7 & Day 1, Group B&51/F & Low vision&Sometimes &Y&Y\\
\hline
F8 & Day 1, Group B&56/F & Low vision&N&Sometimes&Sometimes\\
\hline
F9 & Day 1, Group C&46/F & Totally blind&With help&Using cane&N\\
\hline
F10 & Day 1, Group C&61/F & Low vision&Sometimes &Depends on distance&Y \\
\hline
F11 & Day 1, Group C&45/M & Low vision&N &N&Not answered \\
\hline
F12 & Day 1, Group C&45/M & Low vision&Depends on color&Depends on light&Y \\
\hline
F13 & Day 2, Group D&44/F & Low vision&Y&N&Not answered \\
\hline
F14 & Day 2, Group D&60/F & Low vision&Sometimes or most times, but it has to be at very close proximity and it’s dependent on the level of light&Sometimes. It depends how close it is and only in daylight.&Sometimes, it depends on the light and if the ground is flat.\\
\hline
F15 & Day 2, Group D&76/M & Low vision&N&N&Y \\
\hline
F16 & Day 2, Group D&24/F & Low vision, peripheral vision loss&Depends&Depends&N \\
\hline
F17 & Day 2, Group E&48/F & Low vision, color blind, peripheral vision loss&N&Not answered& N\\
\hline
F18 & Day 2, Group E&62/M & Low vision, no sight in left eye &Y&Y &Y\\
\hline
F19 & Day 2, Group E&66/M &Low vision&Y&Y&Y \\
\hline
F20 & Day 2, Group E&58/F &Low vision&No. I can see light and shapes&Yes. I can see objects in blurry vision&N \\
\hline
F21 & Day 2, Group F&36/F &Low vision&Y&Sometimes but not in the dark&Y \\
\hline
F22 & Day 2, Group F&40/M &Low vision&Y&Y&Y \\
\hline
F23 & Day 2, Group F&34/M &Blind with some light perception&N&N&Not answered \\
\hline
F24 & Day 2, Group F&65/M &Low vision&Y&Blurry vision&Y \\
\hline

\Xhline{2\arrayrulewidth}
\end{tabular}
\end{table*}

\section{Participant Demographics in the Evaluation} \label{sec:evaluation_demographics}
\begin{table*}[htbp]
\scriptsize
\centering
\caption{Participant demographics in the evaluation.}
\label{tab:demographics_evaluation}
\begin{tabular}{>{\raggedright\arraybackslash}m{0.8cm}>{\raggedright\arraybackslash}m{0.8cm}>{\raggedright\arraybackslash}m{1.2cm}>{\raggedright\arraybackslash}m{2.5cm}>{\raggedright\arraybackslash}m{2cm}>{\raggedright\arraybackslash}m{1.2cm}>{\raggedright\arraybackslash}m{4.5cm}}
% 13.5cm
\Xhline{2\arrayrulewidth}
  \textbf{ID} & \textbf{Formative ID} & \textbf{Age/Gender} &  \colorchange{\textbf{Visual Condition}} &\textbf{How Much They Rely on Vision in Daily Activities} & \textbf{Prior AR Experiences}&\colorchange{\textbf{How do you typically navigate the last 10-50 meters when getting to a destination?}}\\
\Xhline{2\arrayrulewidth}
Pilot1 & F7 & 51/F & Low vision, central vision loss & Always & N & Not asked\\
\hline
Pilot2 & F19 & 66/M & Low vision & Always & Y & Not asked\\
\hline
Pilot3 &F4& 58/F & Blind & Not at all & Y & Not asked\\
\hline
Pilot4 & F23 &35/M & Low vision & Sometimes & Y & Not asked\\
\Xhline{2\arrayrulewidth}
P1 & F3 & 20/F& Low vision, central vision loss&Always & Y &A sighted guide (e.g., friend, family member), asking for verbal directions from people nearby, a general navigation app (e.g., Google Maps, Apple Maps)\\
\hline
P2 & F15 &78/M & Totally blind&Not at all & N & A white cane, a sighted guide\\
\hline
P3 & F16 &25/F & Low vision & Sometimes & N & Asking for verbal directions from people nearby, a real-time remote assistance app (e.g. Be My Eyes Volunteer, Aira Agent)\\
\hline
P4 & F4 &33/F & Low vision, sensitive to light, central vision loss &Sometimes & Y &A sighted guide, asking for verbal directions from people nearby, a general navigation app\\
\hline
P5 &-&35/F & Blind with some light perception &Rarely & N&A white cane, a sighted guide, asking for verbal directions from people nearby, a real-time remote assistance app\\
\hline
P6 & F8 & 56/F &  Low vision & Sometimes & N&A white cane, a sighted guide, asking for verbal directions from people nearby, a real-time remote assistance app\\
\hline
P7 & - &55/M &  Low vision, sensitive to light, weak depth perception, central vision loss, scotoma & Always & Y&A white cane, a sighted guide, asking for verbal directions from people nearby, a real-time remote assistance app\\
\hline
P8 & - &46/M & Totally blind & Not at all & N&A sighted guide, Asking for verbal directions from people nearby, a general navigation app, a BLV or accessibility app I can use independently (e.g., Seeing AI, VoiceVista, Be My Eyes without remote assistance)\\
\hline
P9 & - &80/M & Totally blind & Not at all & N&A white cane, asking for verbal directions from people nearby, a real-time remote assistance app\\
\hline
P10 & F2 &44/M &  Low vision, peripheral vision loss, scotoma, occasionally sensitive to light & Always during daylight, cannot see at night& Y&A white cane, a sighted guide, Asking for verbal directions from people nearby, a real-time remote assistance app, a BLV or accessibility app I can use independently\\
\hline
P11 & - &66/F & Low vision, weak depth perception, peripheral vision loss & Sometimes & Y &A white cane, a sighted guide, Asking for verbal directions from people nearby, a real-time remote assistance app\\
\hline
P12 & F1 &46/F & Low vision, central vision loss & Always & Y&A white cane, A sighted guide, asking for verbal directions from people nearby, a general navigation app, a real-time remote assistance app, a BLV or accessibility app I can use independently\\
\hline
P13 & - &42/F & Blind with some light perception, color blind, no depth perception & Not at all & N&A white cane, asking for verbal directions from people nearby\\
\hline
P14 & - &67/M & Low vision, sensitive to light, color blind, weak depth perception, peripheral vision loss & Always & Y&A white cane, a sighted guide, asking for verbal directions from people nearby\\
\hline
P15 & - &50/F & Low vision & Always & Y&A white cane, a sighted guide, asking for verbal directions from people nearby, a general navigation app. If its really important I might do the journey on an earlier day so I don't get lost on the important day. Also I often arrange someone to meet me at a place I know I can find like a tube station.\\
\hline
P16 & F21 &37/F & Low vision, color blind, peripheral vision loss & Always & N& A sighted guide, asking for verbal directions from people nearby, a general navigation app\\
\hline
P17 & - & 49/M & Blind with some light perception& Just to identify day and night & N&A white cane, asking for verbal directions from people nearby\\
\hline
P18 & - & 52/F & Blind with some light perception&Rarely & N &A white cane, a sighted guide\\

% \bottomrule
\Xhline{2\arrayrulewidth}
\end{tabular}
\end{table*}

\section{Note on the Appendices}
Please note that in addition to the above appendices, this paper has a separate Supplementary Materials document.

%TC:endignore

%TC:endignore

\end{document}

\title[\system{}]{\system{}: Enabling In-Situ Spatial Annotation Authoring to Support Exploration and Navigation for Blind and Low Vision People}
\maketitle

\tableofcontents

%%%%%%%%%%%%%%%%%%%%%%%%%%%%%%%%%%%
%%%%%%%%%%%%%%%%%%%%%%%%%%%%%%%%%%%
%%%%%%%%%%%%%%%%%%%%%%%%%%%%%%%%%%%
\section{Note on the Supplementary Materials}
In addition to the following Supplementary Materials, this paper also has an Appendix, which can be found at the end of the full paper. The Appendix contains the participant demographics tables.

%%%%%%%%%%%%%%%%%%%%%%%%%%%%%%%%%%%
%%%%%%%%%%%%%%%%%%%%%%%%%%%%%%%%%%%
%%%%%%%%%%%%%%%%%%%%%%%%%%%%%%%%%%%
\section{Formative Study: Scenario Design Space}\label{sec:formative-scenario-design-space}

The scenarios we presented to participants in the Formative Study were as follows:

\begin{enumerate}
    \item You’re in a location you know well—e.g. a park near your home—what might you want to place in the park for others (friends, family, anyone at all) to hear, or what might you want to hear from others in that location?
    \item You’re at home organizing your room—e.g. organizing a bookshelf, your closet, a collection you have, etc.—would it be helpful to place audio notes in your room? What might you want to place in that space, or what might you want others to place in that space, if anything at all?
    \item You’re travelling to a friend’s apartment complex to house-sit for them while they’re away.
    \begin{enumerate}
        \item Let’s say either the apartment complex itself or your friend has placed notes indoors to help you find the right apartment. Would you imagine this being helpful? What kind of notes would you find helpful?
        \item Once you find the apartment, what kind of notes might you want to hear from your friend in their apartment? These could be related to house-sitting, or not.
    \end{enumerate}
    \item What audio notes would you add to a space if you were telling a story to a young person (e.g. your niece/nephew/child). Where would you tell this story?
    \item What audio notes could O\&M teachers add to a space to act as a training ground, and help you learn?
    \item How could you use audio notes to make a game? Where would you do this? Who would you make this for?
    \item You choose! Where might you place notes, or where would you want to hear notes? If you could “follow” other users of the app, who might you want to follow to hear their audio notes?
\end{enumerate}

Table \ref{tab:formative_scenario_design_space} shows our design space for the above scenarios, including whether in the scenario annotations were authored by the participant (`You') or others, whether the location was private, public, familiar or unfamiliar, and whether the activity was a specific task, for leisure, for the participant (`Yourself') or for others.

\begin{table}[H]
\centering
\caption{The design space for the scenarios we presented to participants in the Formative Study. Note that since we left Scenario 7 up to the participants to choose, we leave the categories blank (`-').}
\label{tab:formative_scenario_design_space}
\begin{tblr}{
  width = .85\linewidth,
  colspec = {Q[100]Q[130]Q[143]Q[143]Q[143]Q[143]},
  cells = {c},
  cell{1}{1} = {r=2}{},
  cell{1}{3} = {c=2}{0.355\linewidth},
  cell{1}{5} = {c=2}{0.352\linewidth},
  vline{2} = {1-9}{},
  vline{3} = {1-9}{},
  vline{5} = {1-9}{},
  vline{3} = {1-9}{},
  hline{3} = {-}{},
}
\textbf{Scenarios} & \textbf{Author}       & \textbf{Location }        &                                & \textbf{Activity }      &                                \\
                   & You (Y) or Others (O) & Private (V) or Public (B) & Familiar (F) or Unfamiliar (U) & Task (T) or Leisure (L) & For Yourself (Y) or Others (O) \\
1                  & Y/O                   & B                         & F                              & L                       & Y/O                            \\
2                  & Y/O                   & V                         & F                              & T                       & Y                              \\
3                  & O                     & V/B                       & U                              & T/L                     & Y                              \\
4                  & Y                     & V/B                       & U/F                            & L                       & O                              \\
5                  & O                     & V/B                       & U/F                            & T                       & Y                              \\
6                  & Y                     & V/B                       & U/F                            & L                       & O                              \\
7                  & -                     & -                         & -                              & -                       & -                              
\end{tblr}
\end{table}

%%%%%%%%%%%%%%%%%%%%%%%%%%%%%%%%%%%
%%%%%%%%%%%%%%%%%%%%%%%%%%%%%%%%%%%
%%%%%%%%%%%%%%%%%%%%%%%%%%%%%%%%%%%
\section{Formative Study: Semi-Structured Discussion Questions}\label{sec:formative-discussion-questions}

Listing \ref{lst:formative-interview-questions} shows the semi-structured discussion questions we asked during the Reflection Activity in the Formative Study.

\begin{lstlisting}[caption={The semi-structured interview questions we asked participants in the formative study.},label={lst:formative-interview-questions},frame=single]
1. After going outside, what were the pain points of 'using the system'?
2. What device would you want to use this on?
3. What surprised you during this workshop? Did you or others come up with any use-cases that you thought were especially interesting/unique or you'd really like to use them for?
4. What features of the system would be most important to include?
5. What features of the system would be least important to include (or which should we definitely not include)?
6. Anything else you want to share with us?
\end{lstlisting}

%%%%%%%%%%%%%%%%%%%%%%%%%%%%%%%%%%%
%%%%%%%%%%%%%%%%%%%%%%%%%%%%%%%%%%%
%%%%%%%%%%%%%%%%%%%%%%%%%%%%%%%%%%%
\section{Formative Study: Annotation Type Quote Percentages}\label{sec:formative_annotation_type_percentages}
There were 140 quotes from the Formative Study identified as annotations. These were categorized into six annotation types (Amenity, Safety, Accessibility, Layout, Attraction, and Experience) through thematic analysis. Some quotes were tagged multiple times (\eg tagged with both layout and accessibility), resulting in a total tag count of 160. We present the number of tags and percent of quotes tagged with the various types in Table \ref{tab:annotation_type_quote_stats}.

\begin{table}[H]
\centering
\caption{The number of times quotes were tagged with annotation types reported as percent of total tags (160) and percent of total identified quotes (140).}
\label{tab:annotation_type_quote_stats}
\begin{tblr}{
  width = \linewidth,
  colspec = {Q[213]Q[108]Q[96]Q[158]Q[96]Q[127]Q[140]},
  vline{2} = {-}{},
  hline{2} = {-}{},
}
& \textbf{Amenity} & \textbf{Safety}  & \textbf{Accessibility} & \textbf{Layout}  & \textbf{Attraction} & \textbf{Experience} \\
\textbf{Tag Count}        & 33               & 30               & 23                     & 70               & 3                   & 10                  \\
\textbf{Percent of Tags}  & 19.53\%          & 17.75\%          & 13.61\%                & 41.42\%          & 1.78\%              & 5.92\%              \\
\textbf{Percent of Quotes} & \textbf{23.57\%} & \textbf{21.43\%} & \textbf{16.43\%}       & \textbf{50.00\%} & \textbf{2.14\%}     & \textbf{7.14\%}     
\end{tblr}
\end{table}

%%%%%%%%%%%%%%%%%%%%%%%%%%%%%%%%%%%
%%%%%%%%%%%%%%%%%%%%%%%%%%%%%%%%%%%
%%%%%%%%%%%%%%%%%%%%%%%%%%%%%%%%%%%
\section{\system{}'s System Prompt}\label{sec:system-prompt}

Listing \ref{lst:system-prompt} shows the prompt \system{} uses when responding to a user query.

\begin{lstlisting}[caption={The prompt we provided our agentic system.},label={lst:system-prompt},frame=single]
You are an AI assistant whose purpose is to aid blind and low vision user in understanding the world around them. You are able to use a number of tools to collect and reason about information relevant to the user, both from sensors on their device such as GPS, compass heading, and also from the internet. To fulfill some requests, you may have to pipe multiple tools sequentially. Try to anticipate user needs when possible. Never shorten the values of latitude or longitude, you should always use them at full precision. When communicating latitude and longitude to the user, you should NEVER use the values directly, but always describe them as relative to the user's latitude and longitude. You should rely on your tools to do all but the simplest of tasks, and combine the output of multiple tools together to provide as complete as possible an answer whenever they are relevant.

From the user's perspective, you are an audio-based assistant on their phone. You hear their voice and respond by voice. You are a voice assistant, so disregard markdown notations like hyperlink and asterisks for titles. You should also disregard hyperlinks per se.

Your user is visually impaired, keep that in mind and adjust your response accordingly. But don't emphasize that to the user explicitly. You should:
- Prioritizing environmental understanding: Provide high-level overviews of space, including shape (e.g., rectangular room, square park) and layout (object positioning relative to space, other objects, and the user). For example, "You are in a rectangular room with a round table in the middle that's about 2 meters wide."
- Being mindful of visual details: Ask if they want information about colors or other visual details to personalize their experience.
- Recording visual conditions: Use tools to record details about their visual condition (e.g., totally blind, low vision, central vision loss) if they share it.
- Sourcing information reliably: Obtain environmental details from Layout annotations or other reliable sources.

When you call any agents, you SHOULD ALWAYS include the User ID, which is a str(uuid.uuid4()), in your query. If the agent asks for it, you should call the agent again with User ID to the agent.

You have access to an agent (CustomizationAgent) that is capable of customizing the user's requirements, including voice, directional instructions, unit preferences, visual feedback colors, visual conditions and verbose level.

The way users interact with you is by clicking a button to start talking. They have to click the button every time. Therefore, you should **never end your response with a question**, like "would you like to know more about it?" as they may think they can directly speak to reply. You should instead say something like "let me know if you want to know more" or "please tell me the specific content of the note".

You have access to key object details at a location (SceneGraphsAgent). Don't list numbers of key objects in layout descriptions unless the user tells you to do so. When describing the layout, you should choose the most important details (e.g., entrances & exits, space size, what's in the center) and offer some position descriptions. As each user has individual language habits, you should consider reasonable synonyms for key objects' names. For example, table tennis table may be referred to as "ping pong table" and gate may be referred to as "exit". However, you should still use the standard name in the scene graph. When users ask about their location, include a specific location name if you know. You should know two things about the user's location: latitude and longitude, and the user's x,y,z coordinates in Unity. When it asks for latitude and longitude, you should input them; otherwise you should input unity coordinates. When retrieving key object positions using the user's latitude and longitude, the key objects' positions are provided in Unity coordinates. You should not give the user Unity coordinates directly. Instead, you should do a calculation and provide descriptions. Be aware that, while you may not know it, the user moves all the time. Therefore, for EVERY inquiry, you should assume that the user has moved, and that all prior results related to a specific location aren't valid anymore. Also, the user may refer to different instances of key objects (even the same kind) at different times. For example, the user may ask a question about one particular podium, then move to another one and ask "what about this one," and you should not rely on previous results to answer questions, but rather use relevant tools each time.
            
You have access to an agent (BLVDBAgent) that handles all annotation-related operations. You should hand all annoatation-relevant requests to this agent. Make sure to always pass the user ID to this agent as this ID is necessary for most of the annotation operations.

You have access to internet search (brave search). If the question is about a key object or something generic (i.e., not directly limited to a note or a key object), in addition to searching the notes, you should ALSO use brave search to find answers. This can be, for example, asking for the history of the current location or a key object (e.g., statue).

You also have access to the camera (i.e., what the user is seeing right now). Call it EVERYTIME when needed as it frequently changes, for example, describing the scene or appearance of an object. When the user asks about what's in front of them, you should call both the OpenAIAgent to understand the image and key objects tool (SceneGraphsAgent). Your answer usually should combine the results, especially include OpenAIAgent's responses. If you think key objects' information is unimportant, don't include it.

 You support wayfinding (SceneGraphsAgent). The user can set a destination and you will offer them turn-by-turn instructions. You SHOULD get the position of the destination from the list of key objects or from an annotation. You should know the key objects' and annotations' distances from the user to find the nearest/second nearest/furthest one. You should use tools to access the data every time and do not rely on your previous function return. If the user wants to meet someone, you should search for whether that person has left an annotation somewhere, and if so, ask the user if they want to navigate to the annotation. When the user asks about the key objects, you should be able to find out their relative positions based on their dimensions and center positions. For example, you should know whether a statue is inside a flower bed.
 
For seating, it's a two-step process: guide them to a seating area, then confirm availability. You'll use key object locations to find seats, or help them navigate if none are nearby. Once there, use the camera image to check for space or obstructions, guiding the user to adjust their view if needed before confirming if the seat is free.
\end{lstlisting}

%%%%%%%%%%%%%%%%%%%%%%%%%%%%%%%%%%%
%%%%%%%%%%%%%%%%%%%%%%%%%%%%%%%%%%%
%%%%%%%%%%%%%%%%%%%%%%%%%%%%%%%%%%%
% \newpage % new page so that the full table fits
\section{Pre-set Annotations for the Evaluation} \label{sec:annotations_inspired_by_quotes}

Table \ref{tab:annotation_list} shows the annotations we added to the local square where participants completed the evaluation study. Note that many of the annotations were inspired by general findings about annotation types, rather than a specific quote.

\begin{longtable}{|p{0.5cm}|p{1.5cm}|p{5.9cm}|p{5.9cm}|}
\caption{Annotations inspired by the formative workshop quotes and findings.}
\label{tab:annotation_list}\\

\hline
\textbf{ID} & \textbf{Type} & \textbf{Example Quote} & \textbf{Annotation}
\endfirsthead

\hline
\textbf{ID} & \textbf{Type} & \textbf{Example Quote} & \textbf{Annotation}
\endhead

\hline
1 & Accessibility & {If it's uneven, it'd be good to have something which tells you when there are steps and things like that.} & The ground in this square is not level. It includes uneven stones and grassy areas.  \\ 
\hline

2 & Accessibility & So like {[}F2{]} said, {[}I{]} also like knowing where the toilets are. & There are no public toilets in this square, but there are some at the {[}nearby{]} station, including one accessible toilet. \\ \hline
3 & Accessibility &  & Visitors must adhere to rules, including no ball games, cycling, or alcohol, and keeping dogs on a leash. \\ \hline
4 & Accessibility & I think personally for me, opening times and closing times, it would have been good to know. Have a note of that, of the different buildings of the whole attraction. & {[}The local{]} square is open seven days a week, from 8am until at least 8pm in the summer. \\ \hline
5-6 & Accessibility & You've got steps coming down.  They're very wide steps or small steps coming down. & Watch out for the stairs. There are four steps. (Attached to two sets of stairs) \\ \hline
7 & Accessibility & I would like to know if there is cracked pavement because that's something that we all struggle with and it's very unsafe for trips and falls right. & Here is a 4-meter-long ramp. Watch out for a small hole near the bottom to avoid tripping! \\ \hline
8 & Amenities & And for me, food is really important. So where the cafes are {[}important{]}. & Across the street, there’s a café with arguably the best flat white in town! \\ \hline
9 & Amenities &  & This notice board advertises various local community programs and support services in {[}the area{]}, covering outdoor activities, health and wellbeing, and cost of living assistance. \\ \hline
10 & Amenities & And the second thing is like a tour guide, but an audio version. & Just south, there's a block that has retail and leisure spots. \\ \hline
11 & Amenities & {[}I would like to know{]} if there's a bus stop and what buses stop there. & The closest bus stop is at {[}the local toy store{]}, just west of the square. \\ \hline
12 & Attraction &  & These white and pink roses are a gift to {[}the city from Bulgarian locals{]}. \\ \hline
13 & Attraction & {If we go to a place of interest, what I would expect the app to do is to make to take me to the place of attraction and to get all the details about the place of attraction.} & The square has a rich history, evolving from an aristocratic residential area to a trade hub, finally becoming a public garden after WWII. \\ \hline
14 & Attraction &  & There are yellow Coreopsis, pink Osteospermum and purple Salvia in the flower bed. \\ \hline

\
15 & Attraction & When I went there, it would have been nice if there was like audio notes of the history. So I had to tell fortunately there was a tour guide but then I was thinking what if he wasn't there and then he left early. I was like, no, I want to know more about the history. & This statue was accidentally won at an auction, when someone waved to greet a friend. He ended up giving the statue as a gift to {[}this square{]}. \\ \hline
16 & Attraction & {The history and the context of it and all the rest.} & There's confusion about whether this statue represents {[}a king from the early or late 1600s{]}. \\ \hline
17 & Attraction & {It could be about nature because this is a park so you can say that it's about like tree species or something like that.} & There are yellow Coreopsis, pink Osteospermum and purple Salvia in the flower bed. \\ \hline
18 & Experience & So it would be nice to have a note, like, `Oh, I came here with my friend, and we went to this building.' & We used the flowerbeds as a hiding spot for stolen biscuits until the ants discovered our stash. \\ \hline
19 & Experience & {It was good to reminisce about, you know, it was a nice happy childhood memory actually.} & I once declared myself “King of the Square” from the podium and was dethroned five minutes later by a well-aimed pinecone. \\ \hline
20 & Experience & I need [an annotation] because... We do a lot of work here and this is where I want them to meet me here. & Hi, this is Ben! Meet me here! \\ \hline
21 & Experience &  & Hi, this is Emma! Meet me here! \\ \hline
22 & Layout & So it's got to clearly define what someone would expect to discover when they first [arrived].  So let's say it's a building with shops either side of the main entrance. & This square has grassy areas on the outside edges and a higher platform in the center. \\ \hline
23 & Layout & Two large fountains with pools of water close by but surrounded by thick walls. & This is the one of the two podiums with a vase on it. All the others here are bare. \\ \hline
24 & Layout &  & This is one of two podiums with a vase, and the only one containing succulents. \\ \hline
25 & Layout & There's a river beneath us... There are poles kind of next to us. & The two historical notice boards in this square are identical, talking about the history and rules. \\ \hline
26 & Layout & Things I would like to be able to know what surround me is like if there is a zebra crossing on that street, so I can go there to cross the road safely. & There aren’t any cross-walks at any of the entrances, but the streets are generally quiet. \\ \hline
27 & Layout & {Let's say you come across a structure or statue {[}...{]} describe the landmark within the park.} & Two identical flower beds sit to the left and right of a central flower bed with a statue in it. \\ \hline
28 & Safety & {It can be quite dangerous also, especially when you're on the pavement and suddenly you fall down on the side of the road.} & Watch out—there’s an elevated flower bed with stone edges and concave cutouts at the corners. Be careful not to trip! \\ \hline
29 & Safety &  & Watch out for the podium! It's close to the flower bed’s edge. \\ \hline
30-34 & Safety &  & Watch out, there’s a podium! (Attached to five different podiums) \\ \hline
35 & Safety & Where the flower beds are...  Please be careful not to cut your hands on the sharp thorns.  & Watch out for the elevated square-shaped flower bed in the center! Be careful not to trip. \\ \hline
36 & Safety &  & Watch out for the table tennis table! People love to play ping pong here. \\ \hline
37 & Safety &  & Watch out for the table tennis table! Many parents bring their kids here. \\ \hline
38 & Safety & Don't fall into the water fountain. & Watch out—there’s an elevated flower bed with stone edges and concave cutouts at the corners. Be careful not to trip! \\ \hline
39 & Safety & I am six foot two and a half and sometimes when I'm walking, even I've got my cane to guide me to avoid objects on the ground, like a tree branch hanging down, it sometimes hits me in the face. So if people who walk past that left a audio note to say, you know, tree branches are quite low on this road, and when I'm coming up to it, I was aware, then i would know when i need to duck. But I get that could change over time if every now and again the local authority cut it. But if they don't, that kind of information would be useful. & Watch out for tree branches hanging from the potted plants here! \\
\hline

\end{longtable}

%%%%%%%%%%%%%%%%%%%%%%%%%%%%%%%%%%%
%%%%%%%%%%%%%%%%%%%%%%%%%%%%%%%%%%%
%%%%%%%%%%%%%%%%%%%%%%%%%%%%%%%%%%%
\section{Evaluation Study Questionnaire and Semi-Structured Interview Questions} \label{sec:evaluation-questionnaire-interview-questions}

Listing \ref{lst:evaluation-questionnaire} shows the questionnaire from the Evaluation Study, and Listing \ref{lst:evaluation-interview-questions} shows the semi-structured interview questions.

\begin{lstlisting}[caption={The questionnaire we asked participants to respond to after the navigation task, including subjective ratings on a 7-point Likert Scale.},label={lst:evaluation-questionnaire},frame=single]
1. Can you describe the route we just navigated? [If low-vision: Please don't look back at the route, just try your best to remember it. Experimenter can stand to block view.]
    1.1 Can you describe the route in general?
    1.2 What landmarks or objects do you recall from the route? [If they described landmarks in prev. Q: Do you remember any other landmarks/objects]
    1.3 Considering each segment in the route, i.e. where you turned during the route,
    1.4 Can you tell me in which direction (e.g. left or right, 3 o'clock etc. if you are facing away from the ramp) each segment went?
    1.5 Can you give your best guess at the distance of each segment (or at least, which parts were longer vs. shorter)?
    1.6 What landmarks or objects do you remember on each segment?
2. Did you learn anything else about the area while you were on the route?
3. On a scale of 1 to 7, where 1 means "strongly disagree" and 7 means "strongly agree."
    3.1 On a scale of 1 to 7: ``This system's capabilities meet my requirements for navigation''
        What about the system met or did not meet your requirements?
    3.2 On a scale of 1 to 7: ``This system is easy to use for navigation''
        Why do you feel this way?
    3.3 On a scale of 1 to 7: The navigation task was very mentally demanding. (Mental Demand)
        Why do you feel this way?
    3.4 On a scale of 1 to 7: The navigation task was very physically demanding. (Physical Demand)
        Why do you feel this way?
    3.5 On a scale of 1 to 7: I accomplished the navigation task I was asked to do very successfully. (Own Performance)
        Why do you feel this way?
    3.6 On a scale of 1 to 7: I felt very insecure, discouraged, irritated, stressed, and annoyed during this navigation task. (Frustration Level)
        Why do you feel this way?
4. Did you encounter any problems along the way? What would you like to change about the system, if anything?
5. What did you like about the system, if anything?
6. What do you think of the amount of information? How relevant/useful was the information?
7. [after both systems] Now that you've experienced both systems, let's compare them. From what perspectives do you find one system more effective or preferable than the other, and conversely, where does the other system stand out?

\end{lstlisting}

\begin{lstlisting}[caption={The semi-structured interview questions we asked participants during the evalutaiton study.},label={lst:evaluation-interview-questions},frame=single]
1. What audio notes did you create, what was their purpose, and who were they for?
2. You created audio notes about this square. When outside the square, what notes would you create, what would be their purpose, and who are they for? It could be anywhere.
3. Did you edit or delete any audio notes? How was that process?
    [If did not use the function]
    Our system also supports editing or deleting audio notes. What do you think of the functions? And, we use natural language for these processes, what do you think of this interaction process?
4. What are some things you learned about the area while using the system?
5. What did you think about the audio notes in the area? What was most useful, what was most interesting, what did you want to filter out, etc.?
6. Did you filter any of the types of notes?
7. What other types of audio notes would you want, if any? [explain all six types of notes]
8. Did you customize the system in any way (e.g. voice, verbosity, colors, etc.)? If so, how?
9. What did you like best about the system?
10. What did you like least about the system- or what did you want to change or add to the system (if anything)?
11. What do you think of the ``form factor'' of the device-a phone attached to your chest with a vest? Would you use something similar in real life? How might you improve it?
12. Assuming we debugged our app and it worked a lot smoother, what could you see yourself using it for?
13. Would you come here alone and explore this square (or somewhere similar) with our app? Why or why not?
14. Anything else you would like to share?

\end{lstlisting}

% %%%%%%%%%%%%%%%%%%%%%%%%%%%%%%%%%%%
% %%%%%%%%%%%%%%%%%%%%%%%%%%%%%%%%%%%
% %%%%%%%%%%%%%%%%%%%%%%%%%%%%%%%%%%%
% \section{TODO: independent localization results}\label{sec:independent-localization-results}

\section{Design Considerations for Interaction and Form Factor}\label{sec:form-factor-design-considerations}
To explore how \system{} might fit into daily routines, we also examined participants’ perspectives on interaction modalities and device form factors and provide more details below.

\subsection{Interaction Initiation} In our experiment, users were required to press a button each time they wanted to interact with the system. While nine participants appreciated that the press-to-speak mechanism was easy,
P7, P13, and P16 suggested that a single press to initiate the back-and-forth interaction flow would be more convenient. As P13 explained, repeatedly pressing the button to ask and respond to questions felt conspicuous, and ``what you don't want to do is highlight the fact that you have something [expensive]''. P13 and P14 also suggested using wake words to initiate interaction with the system.

\subsection{Earphones} We used bone conduction earphones for the majority of participants, with wired earphones provided as alternatives. P9, P13, and P18 appreciated that bone conduction earphones allowed them to hear both the environment and \system{}’s responses and instructions. P13 and P17 also valued their social acceptability. However, P11 and P14 reported occasional conflicts between the bone conduction earphones and their glasses, and P1, P3, and P4 chose wired earphones for personal reasons. These findings suggest that alternative earphone options should always be provided.

%TC:ignore
\begin{figure}[htb]
    \centering
    \includegraphics[width=.5\textwidth]{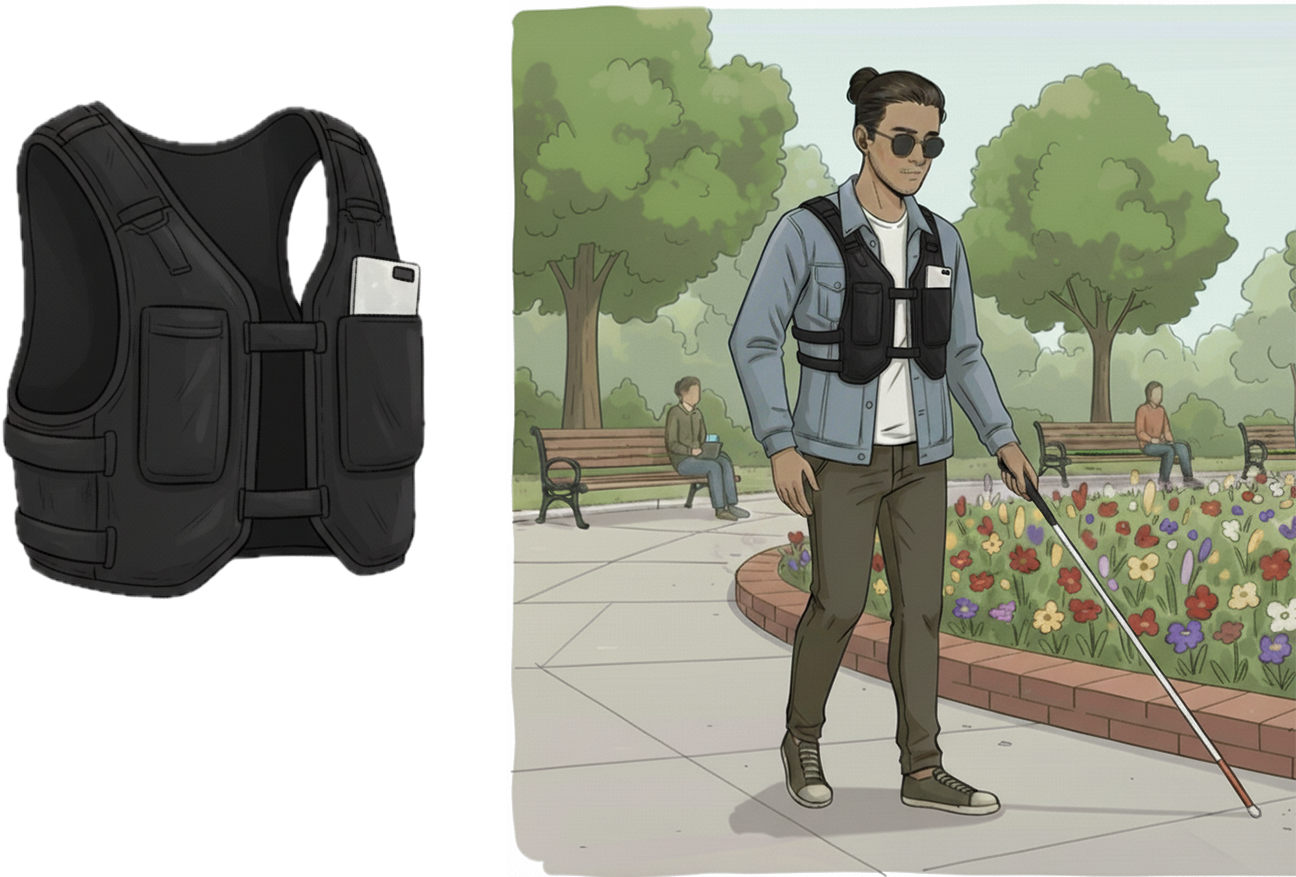}
    \caption{During the study, participants wore a running vest, which held the phone with the camera facing forward. This allowed them to use \system{} hands-free.}
    \Description{Left: An illustration of a black vest with a front pocket. A phone is held in the pocket, with the camera facing forward. Right: An illustration of a man wearing the phone-vest. He is holding a white cane in one of his hands, following the edge of a flowerbed.}
    \label{fig:form-factor}
\end{figure}
%TC:endignore

\subsection{Front-Mounted Phone Design} We adopted a vest with Velcro straps to secure the phone in front of the user for hands-free operation in our experiment, as illustrated in Figure \ref{fig:form-factor}. While participants appreciated hands-free use, twelve participants raised concerns about the phone being exposed to potential thieves, and only two participants (P1, P10) mentioned considering the vest to be safer than holding the phone. Participants also reported that the vest may be cumbersome for daily use (P1, P6, P9) and socially uncomfortable (5/18). P11 and P18 noted that they sometimes unintentionally blocked the camera while the phone was held in front of them.

Four participants (P2, P4, P6, P13) suggested a phone lanyard could make wearing and taking out the phone easier. However, P4 and P17 pointed out that lanyards can shift, reducing the system's accuracy. Eight participants 
suggested smaller attachable cameras instead of relying on the phone camera, whereas P6 and P12 mentioned preferring to keep all functions within a single device. 
P3 and P17 emphasized any additional devices should be low cost. 

While P13 preferred phone-based solutions for portability, five participants recommended smart glasses, as they are hands-free (P14), socially acceptable (P7, P10), and reduce safety risks (P8, P10). In summary, there are varying opinions and trade-offs with respect to providing a hands-free solution that is safe, socially acceptable, and cost-effective.

% \newpage
% \bibliographystyle{ACM-Reference-Format}
% \bibliography{sections/references}